\documentclass[reprint, amsmath, amssymb, aps, prc, longbibliography]{revtex4-2}
\usepackage{dcolumn} 
\usepackage{bm} 
\usepackage{hyperref}
\hypersetup{
    colorlinks=true, 
    linktoc=all,     
    linkcolor=blue,  
}
\usepackage{graphicx} 
\usepackage{multirow}
\usepackage[caption=false, subrefformat=parens, labelformat=empty]{subfig}
\graphicspath{Fig/}
\begin{document}
\title{
    Cold Nuclear Matter Effects on Inclusive $J/\psi$ Production in $p+\text{Au}$ Collisions \newline at $\sqrt{s_\text{NN}}$ = 200 GeV with the STAR Experiment
}
\author{The STAR Collaboration}
\begin{abstract}
    In this paper, a study of cold nuclear matter (CNM) effects is reported based on the new STAR measurement of inclusive $J/\psi$ production in $p+p$ and $p+\text{Au}$ collisions at $\sqrt{s_\text{NN}}$ = 200 GeV, and a combined $J/\psi\rightarrow e^{+}e^{-}$ cross section in $p+p$ collisions at $\sqrt{s}$ = 200 GeV is provided. Given the long-established presumption that the energy density and the temperature produced in proton-nucleon collisions are insufficient to form QGP droplets, CNM effects in $p+\text{Au}$ collisions are quantified by the nuclear modification factor ($R_{p\text{Au}}$), defined as the ratio of the yield of the inclusive $J/\psi$ in $p+\text{Au}$ collisions to that in $p+p$ collisions, scaled by the average number of binary nucleon-nucleon collisions. The $R_{p\text{Au}}$ is derived as a function of transverse momentum ($p_\text{T}$) in the range 4--12 GeV/$c$ and is averaged within the rapidity ($y$) and azimuthal angle ($\varphi$) coverage of $|y|<1$, $0 \leq \varphi < 2\pi$. The result is consistent with unity, suggesting negligible modification of the yield by CNM effects in this kinematic region. Various model calculations are in agreement with the $R_{p\text{Au}}$ measurement, yet the same calculations are less satisfactory in describing the individual invariant yields from $p+p$ and $p+\text{Au}$ collisions. In the covered kinematic region, this analysis has improved the precision of cross-section and invariant-yield measurements in $p+p$ and $p+\text{Au}$ collisions, respectively, and consequently the $R_{p\text{Au}}$.
\end{abstract}
\maketitle
\section{Introduction}
    \indent The Quark-Gluon Plasma (QGP) is a state of matter predicted by Quantum Chromodynamics (QCD) in which quarks and gluons are no longer confined inside hadrons. The QGP is believed to have existed in the early evolution stage of the universe. Great experimental progress has been made in understanding the properties of the QGP through its creation in ultra-relativistic heavy-ion collisions at the Relativistic Heavy Ion Collider (RHIC) at Brookhaven National Laboratory (BNL) and the Large Hadron Collider (LHC) at CERN. \\
    \indent A quarkonium is a bound state of a heavy quark-antiquark pair (${Q\bar{Q}}$). Due to their large masses, heavy quarks are predominantly produced in the early stages of heavy-ion collisions, prior to the formation of the QGP, and are sensitive to the medium throughout the evolution. The binding potentials of different quarkonium have been predicted to be suppressed in the presence of the deconfined medium at different temperatures, which is often viewed as an important signature of the existence of the QGP, making quarkonia unique probes of it. The $J/\psi$ meson is the most abundant quarkonium produced in heavy-ion collisions. The real part of its potential can be color-screened, while the imaginary part is related to the dissociation of $J/\psi$ due to scattering against medium constituents [\onlinecite{MATSUI1986416, PhysRevC.53.3051, PhysRevC.87.044905, Andronic2016, PhysRevLett.99.211602, PhysRevD.100.014008}]. The suppression of $J/\psi$ production has been measured at the Super Proton Synchrotron (SPS) [\onlinecite{ABREU200185}], RHIC [\onlinecite{PhysRevLett.98.232301}], and the LHC [\onlinecite{2011294, Chatrchyan2012, PhysRevLett.109.072301}].\\
    \indent Different physics mechanisms may contribute to such experimental observations. As opposed to those effects related to the hot medium, the manifestations of non-trivial QCD that modify the $J/\psi$ production observed in proton(deuteron)-to-heavy-nucleus collisions are often referred to as the Cold Nuclear Matter (CNM) effects, under the long-established presumption that the energy density and temperature produced in such smaller collision systems are insufficient to form QGP droplets. The nuclear parton distribution function (nPDF) is expected to be modified compared to the free-nucleon PDF, which results in suppression (shadowing) at small Bjorken-$x$ ($x\lesssim0.01$) and enhancement (anti-shadowing) at intermediate $x$ [\onlinecite{PhysRevD.93.085037, Eskola2017}]. The Color Glass Condensate (CGC) effective theory is an alternative framework for accounting for small-$x$ nPDF effects and also provides explanatory power for recombination [\onlinecite{CGC2010}]. In addition, color-octet $c\bar{c}$ pairs may experience energy loss passing through the cold nuclear medium before forming a bound state [\onlinecite{Arleo2013}]. After formation, the $J/\psi$ meson may be broken up by interactions with nucleons in the remnants of the nucleus [\onlinecite{VOGT1999197}] or with co-moving hadrons [\onlinecite{FERREIRO201598}].\\
    \indent Studies of CNM effects with the help of measured spectra and their ratios are essential to solidify these theoretical conclusions. Several measurements [\onlinecite{PhysRevC.87.034904, PhysRevC.102.014902, 2022136865, Acharya2018, CMS:2017exb, c9wp-5tq3}] have been conducted with various collision species in order to study CNM effects. The results indicate that distinct effects are evident across different kinematic ranges. At RHIC energies, ratios of the $J/\psi$ yield in $p+\text{Au}$ and $d+\text{Au}$ collisions relative to $p+p$ collisions are generally flat around unity when the transverse momentum ($p_\text{T}$) is greater than 4 GeV/$c$ in the midrapidity region, while a noticeable suppression is observed in the lower $p_\text{T}$ range [\onlinecite{PhysRevC.87.034904, PhysRevC.102.014902, 2022136865}]. At both RHIC and LHC energies, the forward and backward regions have very different levels of modification compared to those at midrapidity [\onlinecite{PhysRevC.87.034904, PhysRevC.102.014902, Acharya2018}]. There is also the possibility of the formation of a QGP droplet in such small collision systems, causing the dissociation of $J/\psi$ [\onlinecite{DU2015147, Du2019}].\\
    \indent The cross section of inclusive $J/\psi$ in $p+p$ collisions is needed in order to quantify CNM effects. Measurement of the $p+p$ cross section also provides constraints on model calculations and could shed light on the production mechanism of the $J/\psi$ meson. The production of $J/\psi$ in $p+p$ collisions can be factorized into soft and hard processes, which are associated with long and short distance interactions, respectively [\onlinecite{PhysRevD.74.074007}]. The hard processes reflect the production of $c\bar{c}$ from hard parton scatterings and can be estimated by perturbative Quantum Chromodynamics (pQCD), while the soft processes account for the evolution from a $c\bar{c}$ pair into bound states, which is usually parameterized by phenomenological models [\onlinecite{Andronic2016}]. \\
    \indent In this paper, a new measurement of the production of inclusive $J/\psi$ at $\sqrt{s_\text{NN}} = 200\text{~GeV}$ in $p+\text{Au}$ collisions, as well as an improved measurement of the inclusive $J/\psi\rightarrow e^{+}e^{-}$ cross section in $p+p$ collisions, are reported. The CNM effects on the production of inclusive $J/\psi$ are quantified by the nuclear modification factor ($R_{p\text{Au}}$). The improved cross section in $p+p$ is a combination of published STAR [\onlinecite{201355, 201887}] results and the present measurement.
\section{Experiment}
    \indent Located at BNL, RHIC is a versatile accelerator dedicated to studying ion-ion collisions, the QGP, and the structure of the proton. It is capable of performing relativistic collisions between different species of heavy ions at various center-of-mass energies. The data used in the new measurement were collected by STAR in 2015. The reconstruction of $J/psi$ at mid-rapidity relies on tracking and calorimetry provided by several STAR subsystems. The Time Projection Chamber (TPC) is centered around the beam pipe with a pseudorapidty coverage of $|\eta|<1.8$. It serves as the main tracking detector [\onlinecite{ANDERSON2003659}], while providing particle identification (PID) through measurements of ionization energy loss ($\text{d}E/\text{d}x$). The Barrel Electromagnetic Calorimeter (BEMC) detector can provide complementary PID for charged particles with mid-to-high momentum by measuring energy deposition in its materials. In addition, each event is triggered by the high-tower (BHT2) trigger, which selects events with particles that have high transverse energy ($E_\text{T}$), by requiring energy deposition in at least one of the towers to exceed a certain threshold [\onlinecite{BEDDO2003725}]. The acceptance of the BEMC spans $|\eta|<1$. The scintillator-based Beam-Beam Counter (BBC) [\onlinecite{10.1063/1.2888113}] provides the prerequisite BBC Minimum-Bias (BBCMB) trigger for the BHT2 trigger, by requiring a coincidence signal between the east and west modules of the BBC. The BBC covers a pseudorapidity ($\eta$) range of $3.6<|\eta|<5.2$. The BBC, BEMC, and TPC all have full azimuthal acceptance.
\section{Data Analysis}
    \indent The nuclear modification factor $R_{p\text{Au}}$ is defined as:
    \begin{equation}
        \label{eq:RpAu}
        R_{p\text{Au}}=\frac{1}{\langle T_\text{AA}\rangle}\times\frac{\left(\frac{\text{d}^{2}N}{p_\text{T}\text{d}p_\text{T}\text{d}y}\right)_{p+\text{Au}}}{\left(\frac{\text{d}^{2}\sigma}{p_\text{T}\text{d}p_\text{T}\text{d}y}\right)_{p+p}},
    \end{equation}
    where the $\langle T_\text{AA}\rangle = \langle N_\text{coll}\rangle/\sigma_\text{NN}^\text{inel.}$ is the nuclear thickness function calculated using the Glauber model, in which 
     $\langle N_\text{coll}\rangle = 4.7\pm 0.3$ [\onlinecite{PhysRevC.102.014902}] is the average number of binary nucleon-nucleon collisions in minimum bias $p+\text{Au}$ collisions, and $\sigma_\text{NN}^\text{inel.} = 42 \text{~mb}$ [\onlinecite{PhysRevLett.91.172302}] is the inelastic nucleon-nucleon cross section at $\sqrt{s_\text{NN}} = 200 \text{~GeV}$. The expression $\left(\frac{\text{d}^{2}N}{p_\text{T}\text{d}p_\text{T}\text{d}y}\right)_{p+\text{Au}}$ is the invariant yield per unit rapidity in $p+\text{Au}$ collisions, and $\left(\frac{\text{d}^{2}\sigma}{p_\text{T}\text{d}p_\text{T}\text{d}y}\right)_{p+p}$ is the cross section in $p+p$ collisions. \\
    \indent In the following, we describe how the invariant yield in $p+p$ and $p+\text{Au}$ collisions from Equation \ref{eq:RpAu} is determined in each $p_\text{T}$ bin from the data using the same procedure. It is calculated with the following formula:
    \begin{equation}
        \frac{\text{d}^{2}N}{2\pi p_\text{T}dp_\text{T}dy}=
        \frac{1}{2\pi p_\text{T}\Delta p_\text{T}\Delta y}\cdot
        \frac{Trig.Bias}{N_\text{MB}^\text{eqv.}\varepsilon_\text{MB}^\text{vtx.}}\cdot
        \frac{N_{J/\psi}^\text{raw}}{\varepsilon_{J/\psi}^\text{RC}}
    \end{equation}
    where $\Delta p_\text{T}$ is width of the $p_\text{T}$ bin, $\Delta y$ is the width of the rapidity range, $Trig.Bias=\frac{\varepsilon_\text{MB}^\text{BBC}\varepsilon_\text{MB}^\text{vtx.}}{\varepsilon_{J/\psi}^\text{BBC}\varepsilon_{J/\psi}^\text{vtx}}$ accounts for the difference in the BBC trigger efficiency ($\varepsilon^\text{BBC}$) and vertex finding efficiency ($\varepsilon^\text{vtx.}$) between $J/\psi$ events and minimum bias events, $N_\text{MB}^\text{eqv.}$ is the equivalent number of minimum bias events studied, $N_{J/\psi}^\text{raw}$ are the raw counts, and $\varepsilon_{J/\psi}^\text{RC}$ quantifies the acceptance and reconstruction efficiency of $J/\psi$. The efficiencies are extracted from STAR embedding simulations, while the $J/\psi$ signal fit templates used to obtain the raw counts are derived from the EvtGen generator [\onlinecite{LANGE2001152}]. The invariant yield in $p+p$ collisions is multiplied by the non-single-diffractive cross section ($\sigma^\text{NSD}_{pp} = 30.0 \pm 2.4 \text{~mb}$ [\onlinecite{PhysRevD.86.072013}]) to obtain the cross section.
    \subsection{Event Selection, Track Selection and Electron/Positron Identification}
        \indent The events are selected by requiring the distance along the beam direction from the center of the TPC to the reconstructed primary vertex to be less than 80~cm. Such primary vertices are also required to have sufficient support from primary tracks [\onlinecite{Smirnov_2017}].\\
        \indent The TPC provides information to track the electrons and positrons ($e^{\pm}$). To ensure reliable tracking, tracks are required to have at least 20 hits in the TPC for track fitting ($nHitsFit \geq 20$) and at least ten hits in the TPC for $\text{d}E/\text{d}x$ evaluation ($nHitsdEdx \geq 10$). In order to reduce the possibility of selecting duplicate tracks, the ratio of the number of hits used to fit the track to the total number of possible hits along this track is required to be no less than 0.52. To reduce the possibility of selecting tracks associated with a secondary vertex, the Distance of Closest Approach (DCA) to the primary vertex is required to be no more than 1.5~cm. A minimum $p_\text{T}$ of 1 GeV/$c$ is required for the tracks in order to make sure that the charged particles reach the BEMC. Furthermore, the extrapolation of the track to the BEMC radius is required to match with the nearest BEMC cluster in $\eta$ and $\varphi$. To accommodate the BEMC acceptance, all tracks are required to have $|\eta|<1$.\\
        \indent The selected tracks are identified as $e^{\pm}$ candidates if they pass the following two selections simultaneously.
        \subsubsection{PID by TPC with $n\sigma_{e}$}
            \indent $n\sigma_{e}$ is defined as:
            \begin{equation}
                n\sigma_{e}=
                \frac{1}{\sigma_{\text{ln}\left(\frac{\text{d}E}{\text{d}x}\right)_{e}}}\times \text{ln}\frac{\left(\frac{\text{d}E}{\text{d}x}\right)_\text{measured}}{\langle\left(\frac{\text{d}E}{\text{d}x}\right)_{e}\rangle},
            \end{equation}
            where $\left(\frac{\text{d}E}{\text{d}x}\right)_\text{measured}$ is the energy loss measured for a track; the expected energy loss $\langle\left(\frac{\text{d}E}{\text{d}x}\right)_{e}\rangle$ is obtained based on the Bichsel formalism [\onlinecite{BICHSEL2006154}] with the measured momentum when assuming such a track corresponds to $e^{\pm}$; $\sigma_{\text{ln}\left(\frac{\text{d}E}{\text{d}x}\right)_{e}}$ is the resolution of the natural logarithm of the energy loss for electrons and positrons. The nominal selection on $n\sigma_{e}$ is $-1.5 < n\sigma_{e} < 2.5$.
        \subsubsection{PID by BEMC with E/p}
            \indent The ratio ($E/p$) of the energy ($E$) deposited by a particle in the BEMC to the magnitude of the particle's momentum ($p$) measured in the TPC, is used to further select $e^{\pm}$. It is expected to be roughly unity for $e^{\pm}$. The nominal selection on $E/p$ is $0.5 < E/p <2.5$.
    \subsection{Signal Extraction}
        \label{subsec:rawyield}
\begin{figure}[h!]
            \centering
            \subfloat[\label{fig:fig1.a}]{
                \includegraphics[width = 8.2 cm]{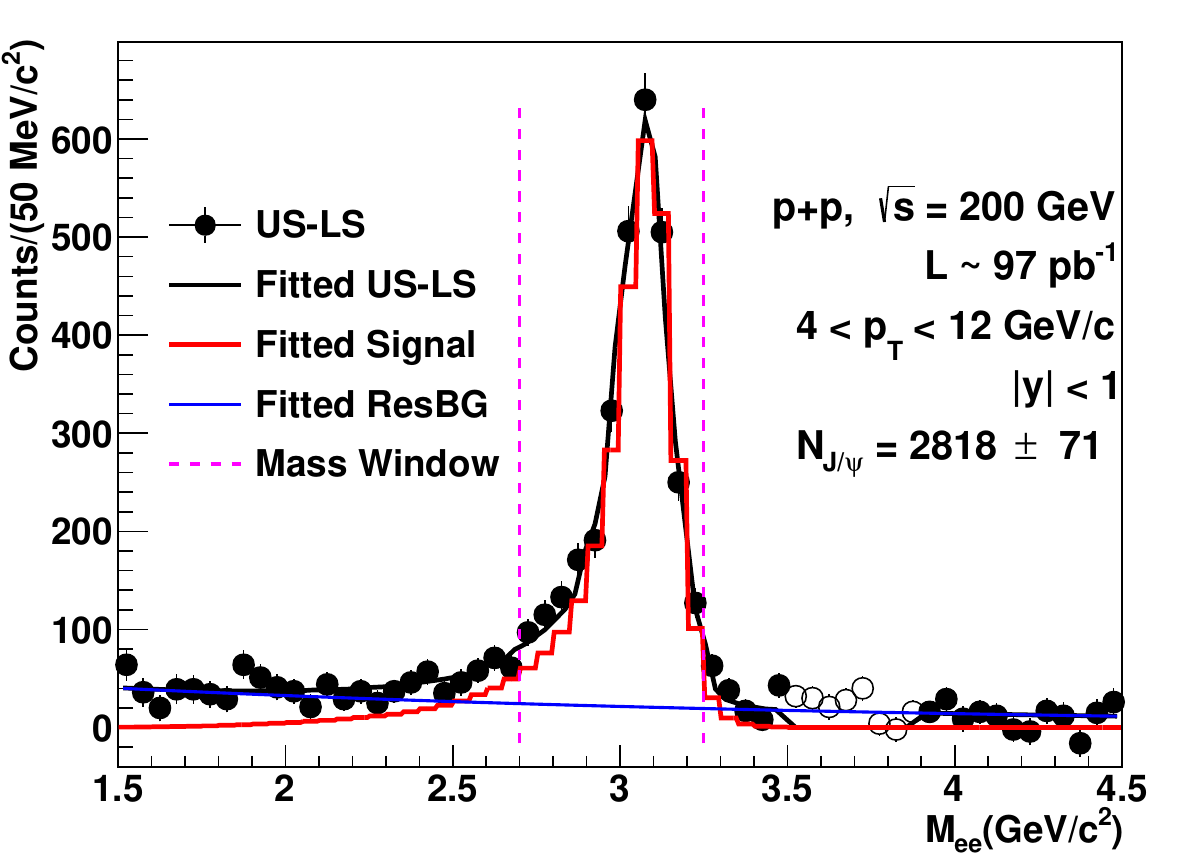}
            }\\
            \subfloat[\label{fig:fig1.b}]{
                \includegraphics[width = 8.2 cm]{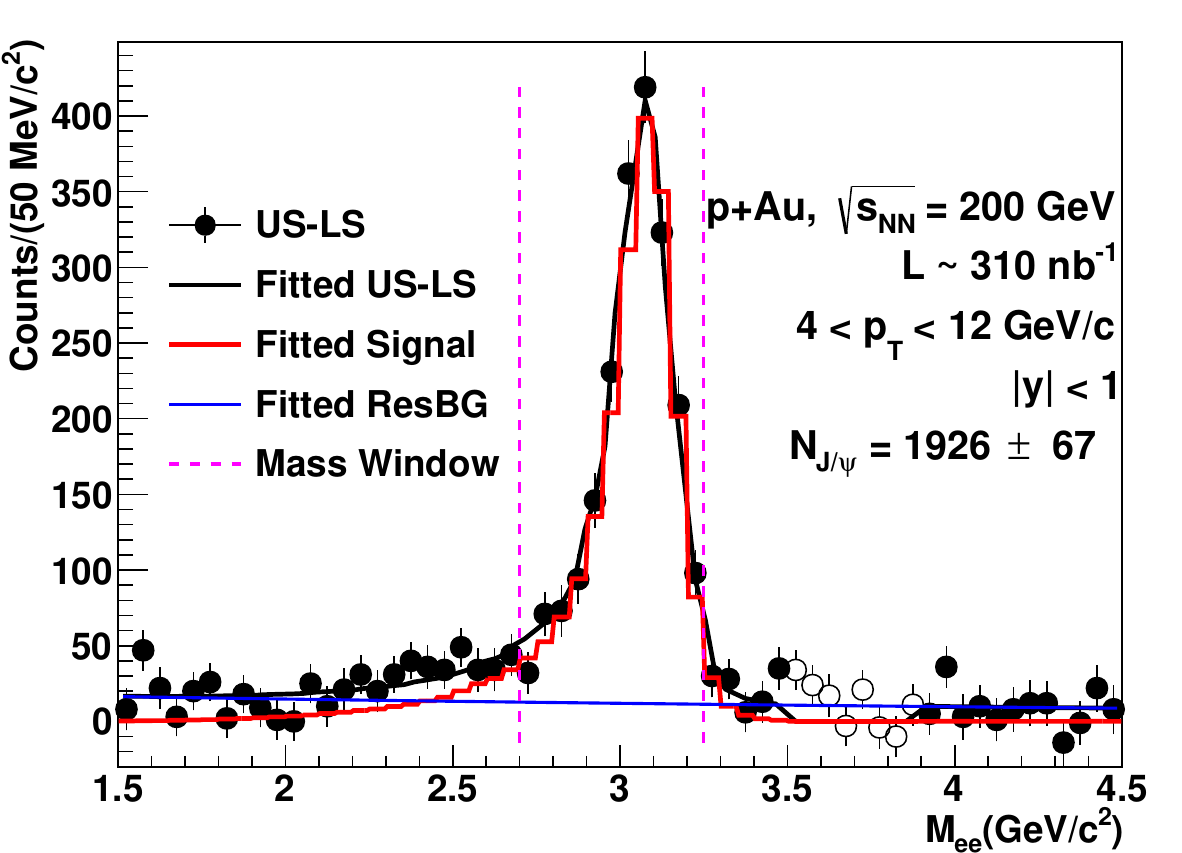}
            }
            \caption{
                \label{fig:fig1_rawyield}
                The invariant mass distributions of $e^{\pm}$ pairs in $p+p$ and $p+\text{Au}$ collisions. The vertical bars represent statistical uncertainties. The open black circles are excluded from the fit to reduce potential influence from the $\psi\left(2S\right)$ meson.
            }
        \end{figure}

        \indent Each $J/\psi$ candidate is reconstructed from an $e^{-}e^{+}$ pair (the unlike-sign pair, $US$) of selected $e^{\pm}$ tracks that have opposite electric charges. For the BHT2 trigger, the online $E_{\text{T}}$ threshold is approximately 4.2 GeV and at least one of the $e^{\pm}$ is required to have $p_{\text{T}}$ greater than 4.3 GeV/$c$. Such a $p_{\text{T}}$ cutoff in the calculation of $R_{p\text{Au}}$ is lowered to 3.5 GeV/$c$ to reduce the statistical uncertainty below the trigger threshold, at the expense of increased systematic uncertainty. This leads to only a limited increase in the systematic uncertainty, because the $p+p$ and $p+\text{Au}$ data were taken in the same year, with the same trigger setup and detector configuration. The combinatorial background is estimated by making the $e^{-}e^{-}$ or $e^{+}e^{+}$ pairs (the like-sign pair, $LS$) with the same electric charge in the same event. After subtracting the combinatorial background, the $US-LS$ invariant mass distribution is fit with the sum of a reconstructed $J/\psi$ mass template plus a function to describe the residual background, as illustrated in Fig.~\ref{fig:fig1_rawyield} for $4<p_\text{T}<12$~GeV/$c$ and $|y|<1$. The reconstructed $J/\psi$ mass template is generated by EvtGen, described in detail in Section \ref{subsec:eff}. The residual background is modeled as an exponential function. The following is an exhaustive list of the fit parameters: 1) a parameter that quantifies the momentum resolution and affects the reconstructed $J/\psi$ invariant mass resolution; 2) the positive scaling factor on the reconstructed $J/\psi$ mass template reflecting the yield; 3) the exponential coefficient in the residual background; 4) the positive scaling factor for the residual background. The raw $J/\psi$ yield is extracted by integrating the $US-LS$ mass distribution and subtracting the integral of the fitted residual background within a nominal mass window of $2.7 < M_{ee} < 3.25 \text{~GeV/}c^{2}$. The loss of tails outside the integral window is accounted for by the efficiency correction.
    \subsection{Acceptance and Efficiency Corrections}
        \label{subsec:eff}
        \indent The $n\sigma_{e}$ PID efficiency is obtained from a study using $e^{\pm}$ pairs from photon conversions and Dalitz decays of $\pi^\text{0}$, $\eta$ and $\eta'$, often called photonic dielectrons. The pair invariant mass is required to be smaller than 0.24 GeV/$c^{2}$, while the distance of the closest approach between the tracks within the pair is limited to 1~cm maximum, in addition to the nominal track level requirements. A photonic $e^{\pm}$ enriched sample is obtained by this selection. The $n\sigma_{e}$ PID efficiency is extracted from the $n\sigma_{e}$ distribution of this sample.\\
        \indent The embedding simulation is used to evaluate the correction factor for the tracking efficiency, the $E/p$ PID efficiency, and the acceptance of the TPC and the BEMC. The products of simulated $J/\psi\rightarrow e^{+}e^{-}$ decay are distributed according to the phase space, providing the Monte Carlo kinematics of the particles of interest. The decays are propagated into a simulation of the detector response using the GEANT3 package with the STAR geometry and configuration. The simulated detector hits are mixed with those in experimental collision data, and reconstruction of the simulated tracks follows the same approach as in the experimental data.\\
        \indent The STAR embedding procedure underestimates the momentum resolution (i.e., overestimates the performance), therefore an additional smearing is applied to the track momentum, by scaling the relative deviation of the reconstructed $p_\text{T}$ from the simulated $p_\text{T}$ of each track. The width of the relative deviation distribution as a function of simulated $p_\text{T}$ is empirically parameterized as $\sigma_{\hat{\Delta} p_\text{T}}\left(p_\text{T};a,b\right)=\sqrt{a^{2}p_\text{T}^{2}+b^{2}}$ and the additional smearing pattern is generated by varying $a$. A similar procedure to smear the reconstructed $p_\text{T}$ from embedding was also used in a previous STAR publication~[\onlinecite{2022136865}] on quarkonia production.\\
        \indent The aforementioned reconstructed $J/\psi$ mass template used in the signal extraction fit is also dependent on the embedding data. It is generated by folding the efficiencies and the additionally smeared momentum resolution of a single $e^{\pm}$ with the EvtGen generator [\onlinecite{LANGE2001152}], thereby accounting for the final-state radiative correction. Since all inputs from the embedding samples depend on $a$, the templates are inherently dependent on $a$. It should be noted that, in the previous subsection, the first parameter in the fit function used to extract the raw yield is $a$.\\
        \indent The equivalent number of minimum bias events $N_\text{MB}^\text{eqv.}$, the BBC trigger efficiency and the vertex finding efficiency in $p+p$ and $p+\text{Au}$ collisions are obtained by studying PYTHIA [\onlinecite{Sjostrand:2006za, SJOSTRAND2008852}] and HIJING [\onlinecite{Buckley2015}] embedding samples, respectively. (See Appendix \ref{apdx:apdx2_HIJING} for detailed settings.) The PYTHIA and HIJING events are mixed into experimental zero-bias data, which have no trigger requirement and are recorded at random times. For $p+p$, samples for minimum bias events are generated, and those containing a $J/\psi$ within $|y|<1$ are selected as $J/\psi$ events. The former is for the purpose of determining $N_\text{MB}^\text{eqv.}$, while the latter is used to calculate the BBC trigger efficiency and vertex finding efficiency. For $p+\text{Au}$, events with $D^{0}$ mesons are selected instead of $J/\psi$ due to the lack of implementation of quarkonia in HIJING. For $p+p$, two generator settings are used: the STAR official heavy flavor tune (see Appendix \ref{apdx:apdx1_HF_TUNE} for detailed settings), and the tune ``4Cx'' [\onlinecite{Corke2011}]. The latter features a Gaussian hadronic-matter profile with $x$-dependent width, on top of the PYTHIA 8 default settings. An additional uncertainty originating from the difference between the tunes is assigned. Due to the high luminosity of the $p+p$ and $p+\text{Au}$ collisions in 2015 at $\sqrt{s_\text{NN}} = 200\text{~GeV}$, the probability of having more than one collision in a single bunch crossing cannot be neglected. A correction for this effect is determined by calculating the probabilities of having different numbers of collisions in the same bunch crossing, with higher-order effects (more than two collisions) ignored. The fraction of having two collisions in a bunch crossing among any collision found in a bunch crossing is estimated to be $0.107\pm0.001$ for $p+p$ and $0.05\pm0.01$ for $p+\text{Au}$. The in-bunch pileup correction is applied to the equivalent number of minimum-bias events and to the BBC trigger efficiency.
    \subsection{Systematic Uncertainties}
        \indent The systematic uncertainty is estimated from the following sources: raw yield estimation, tracking efficiency, TPC efficiency (using $n\sigma_{e}$), BEMC PID efficiency (using $E/p$), BHT2 trigger efficiency, BBC trigger efficiency, and vertex finding efficiency. These sources are assumed to be uncorrelated. The result with the nominal cuts/configuration is taken to be the central value of the measurement, and the difference between the central result and results calculated with varied cuts/configurations is assigned as the systematic uncertainty for that source. The uncertainties for raw yield estimation, tracking efficiency, TPC efficiency, BEMC PID efficiency, and BHT2 trigger efficiency are $p_\text{T}$ dependent, while the uncertainties of the remaining sources are $p_\text{T}$ independent and are quoted as the ``normalization uncertainty.'' Table \ref{table:tab2_syserr} summarizes the systematic uncertainties in this analysis. The evaluation of the three physics observables is shown. The $p_\text{T}$ dependent uncertainties are given as a range.\\
        \indent For the raw yield estimation, the effects of the fit range, mass integral window, residual background formula, mass bin width, integration method, and additional smearing are taken into account. The lower and upper limits of the fit range and mass integral window are varied by $\pm0.05$ independently (8 variations in total). The residual background formula is varied from the exponential function to a linear function. The mass bin width is varied to half the nominal width, from 0.05 GeV/$c^{2}$ to 0.025 GeV/$c^{2}$. The nominal integration method is subtracting the integral of the residual background contribution of the fit function from the integral of the $US-LS$ $e^{\pm}$ pair mass distribution, while the variation is the integral of the signal contribution of the fit function (the template integral scaled by the signal fit normalization). The uncertainty in additional smearing is estimated by varying the parameter $a$ over its 68\% confidence interval. The maximum deviation among the 13 variations from the central value is assigned as the uncertainty contributed by the raw yield calculation.\\
        \indent The uncertainty of the tracking efficiency is estimated by varying the combination of track selection cuts $(nHitsFit, nHitsdEdx, DCA)$ simultaneously. \\
        \indent Finally, $\sigma^\text{NSD}_{pp}$ contributes an 8\% normalization uncertainty to the $p+p$ cross section calculation [\onlinecite{PhysRevD.86.072013}], while in $R_{p\text{Au}}$ the value of $N_\text{coll}$ contributes another 6.4\% normalization uncertainty.
        \begin{table}[h!]
            \caption{{\label{table:tab2_syserr}} Summary table of the systematic uncertainties from this analysis in percentages.}
            \centering
            \begin{ruledtabular}
                \begin{tabular}{cccc}
                    \multicolumn{4}{c}{Systematic Uncertainties in 2015 [\%]} \\
                    Source & $p+p$ & $p+\text{Au}$ & $R_{p\text{Au}}$\\
                    \colrule
                    Raw Yield & 2.8--5.1 & 1.7--4.1 & 1.6--5.7\\
                    Tracking Eff. & $< 3.9$ & $< 3.4$ & $< 2.2$\\
                    TPC PID Eff. & $< 5.4$ & $< 11.2$ & $< 7.1$\\
                    BEMC PID Eff. & $< 2.3$ & $< 4.2$ & $< 4.1$\\
                    BHT2 Trigger Eff. & $< 1.6$ & $< 0.6$ & $< 0.7$\\
                    $N_\text{mb}^\text{eqv.}$*\footnote{The $N_\text{mb}^\text{eqv.}$* is the quadrature sum of contributions from the $N_\text{mb}^\text{eqv.}$, $Trig.Bias$ and vertex finding efficiency estimation, and is dominated by the former 2. } & 3.0 & 5.3 & 6.1\\
                    $\sigma_{pp}^\text{NSD}$ & 8.0 & N/A & 8.0\\
                    $N_\text{coll.}$ & N/A & N/A & 6.4\\
                \end{tabular}
            \end{ruledtabular}
        \end{table}
    \subsection{Combination with measurements in 2009 and 2012 for $p+p$ cross-section}
        \indent The inclusive $J/\psi$ dielectron cross section in $p+p$ collisions from this analysis is combined with published STAR results using data taken in 2009 and 2012. They share the same rapidity and azimuthal coverage, but with different $p_\text{T}$ ranges, although the binning scheme within the overlapping $p_\text{T}$ range is the same. Each $p_\text{T}$ bin is combined independently. The measurements are combined with the \textbf{B}est \textbf{L}inear \textbf{U}nbiased \textbf{E}stimator (BLUE), where ``\textbf{B}est'' stands for the minimized combined variance, ``\textbf{L}inear'' indicates the combination is linear and can be viewed as a weighted sum, and the requirement of having an ``\textbf{U}nbiased \textbf{E}stimator'' results in the sum of the weighting factors being 1 [\onlinecite{Nisius2014}]. Table \ref{table:tab3_err_combine} demonstrates the effectiveness of the combination. The systematic uncertainties from the TPC efficiency and raw yield calculations are assumed to be uncorrelated, given their data-driven nature. The other sources of systematic uncertainty are conservatively assumed to have correlation coefficients equal to 1. The asymmetric uncertainties from the raw yield estimation in the 2009 results were averaged in order to provide a direct comparison with other years. Certain sources were not included in the original analyses, marked by ``-''. In these cases, their uncertainties and the combined cross section are estimated from the analysis in which the uncertainties were evaluated. Additional systematic uncertainty sources exist in the 2012 measurement due to the use of the Time-of-Flight (TOF) detector. The difference in the resulting $J/\psi$ reconstruction efficiency between $J/\psi$ embedding and $e^{\pm}$ embedding (Emd. Type) was treated as a normalization uncertainty in the 2009 analysis, but is corrected for as a systematic bias in the 2015 analysis, noted by \ref{table:tab3_err_combine} \ref{footnote:emdtype}. The total uncertainty listed in the table for each individual measurement accounts for all the sources, including those marked by ``-''. The combination reduces the total uncertainty as expected.\\
        \begin{table}[h!]
            \caption{
                \label{table:tab3_err_combine} Uncertainties of dielectron measurements ($4 < p_\text{T} < 10\text{~GeV/}c$), quoted as the percentage of the central value within the overlapping $p_\text{T}$ range. The $10 < p_\text{T} < 12\text{~GeV/}c$ bin is not included due to the fact that the sum of certain systematic uncertainties was extrapolated from lower $p_\text{T}$ ranges in the 2012 results, inhibiting the comparison.
            }
            \centering
            \begin{ruledtabular}
                \begin{tabular}{ccccc}
                    \multicolumn{5}{c}{Systematic Uncertainties in Different Data Sets [\%]} \\
                    Year & 2009 & 2012 & 2015 & Combined\\
                    \colrule
                    Stat. & 10.2--27.3 & 5.9--20.7 & 5.1--18.9 & 4.0--13.7\\
                    Raw Yield & 4.4--13.0 & 1.2--6.2 & 2.8--5.1 & 1.8--3.6\\
                    Trk. Eff. & - & 2.1--3.4  & Negligible & 0.3--1.8\\
                    TPC Eff. & - & 1.4--4.9 & 0.0--5.4 & 0.6--2.4\\
                    BEMC Eff. & 0.0--14.6 &  0.3--7.5 &  0.0--2.3 & 0.8--2.7\\
                    TOF Eff. & N/A & 3.2--5.5 & N/A & 0.4--2.9\\
                    BHT2 Eff. & 0.4--5.3 & 0.2--0.9 & 0.0--1.6 & 0.1--1.0\\
                    $\sigma_{pp}^\text{NSD}$ & 8.0 & 8.0 & 8.0 & 8.0\\
                    $N_\text{mb}^\text{eqv.}$ & - & 3.6  & 3.0 & 3.1--3.5\\
                    Emd. Type & 7.5 & - & -\footnote{\label{footnote:emdtype}Treated as correction and contributes to statistical uncertainty} & 0.8--5.4\\
                    Total & 16.7--32.4 & 13.8--19.9 & 10.9--18.6 & 10.4--12.7 \\
                \end{tabular}
            \end{ruledtabular}
        \end{table}
        \indent An attempt is also made to use the combined $p+p$ result as the reference measurement in the calculation of $R_{p\text{Au}}$. In this potential $R_{p\text{Au}}$ calculation, due to the inclusion of 2009 and 2012 $p+p$ data, the trigger $e^{\pm}$ $p_\text{T}$ requirement for both $p+p$ and $p+\text{Au}$ data taken in 2015 can only be set to 4.3 GeV/$c$ instead of the treatment mentioned in Section \ref{subsec:rawyield}. The $R_{p\text{Au}}$ calculation with the combined $p+p$ reference does not improve upon the calculation using only the 2015 $p+p$ reference (with a 3.5 GeV trigger $p_\text{T}$ cutoff), and is therefore discarded.
\section{Results and Discussions}
\begin{figure}[h!]
        \subfloat[\label{fig:fig2.a}]{
            \includegraphics[width = 8.2 cm]{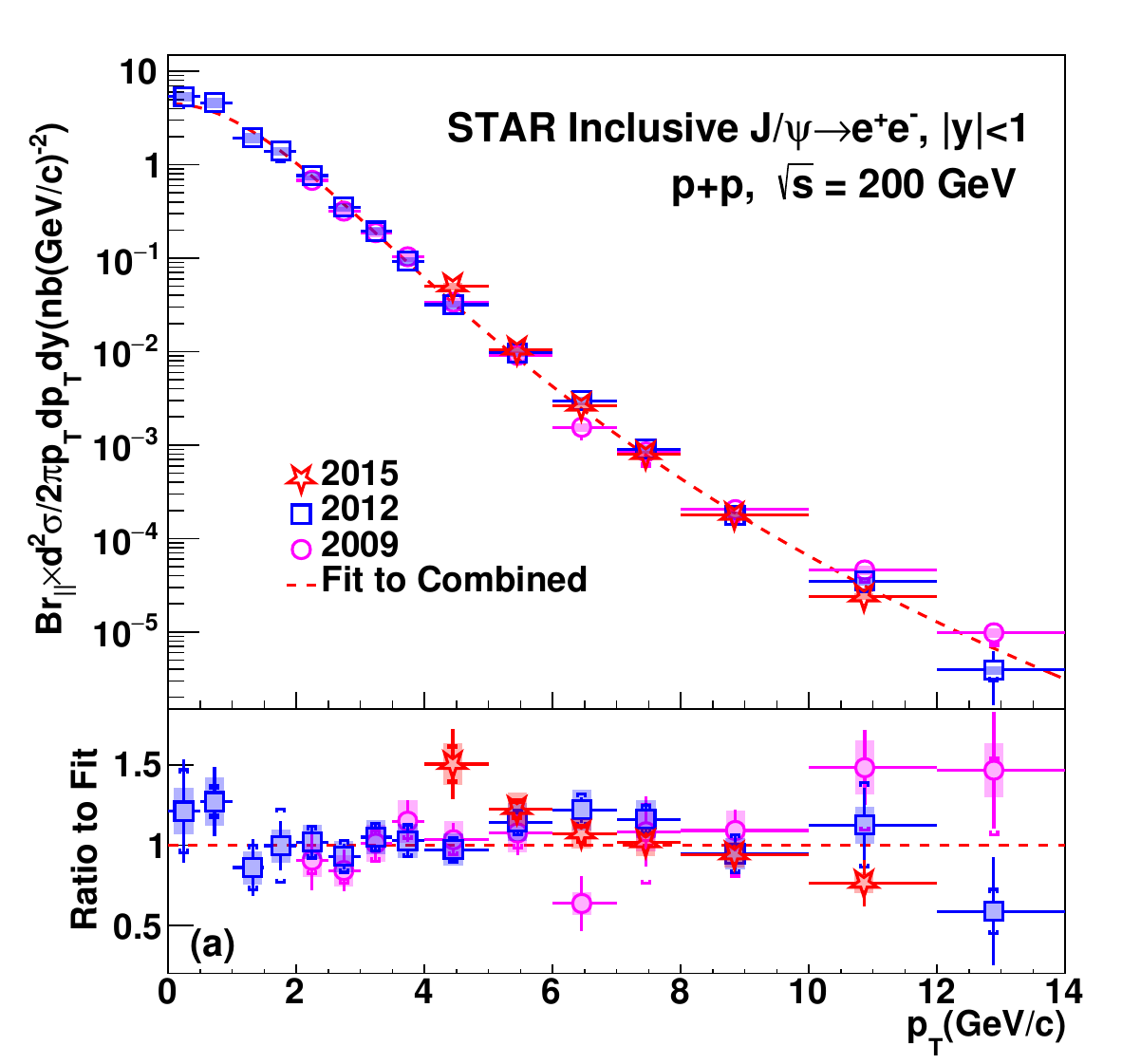}
        }\\
        \subfloat[\label{fig:fig2.b}]{
            \includegraphics[width = 8.2 cm]{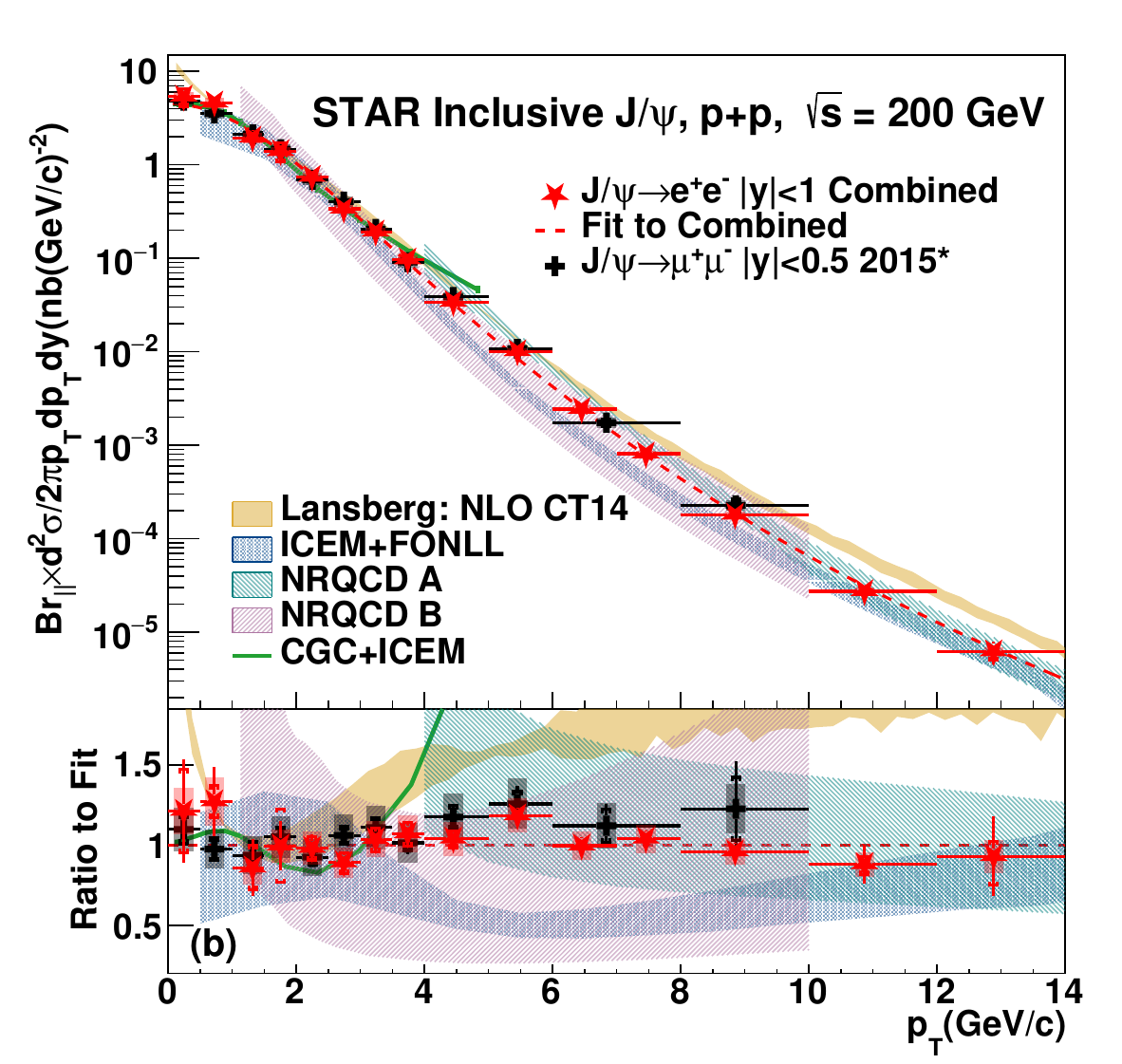}
        }
        \caption{
            \label{fig:fig2_pp_crosssec} Inclusive $J/\psi\rightarrow e^{+}e^{-}$ cross section as a function of $p_\text{T}$ in $p+p$ collisions at $\sqrt{s} = 200\text{~GeV}$. (a) Combined results using data taken in 2009, 2012, and 2015. (b) The combined cross section being compared to the STAR $J/\psi\rightarrow \mu^{+}\mu^{-}$ measurement for $|y| < 0.5$ at the same $\sqrt{s_\text{NN}}$, and to various model calculations. The ``*'' at the year of the dimuon measurement indicates that it is corrected for the rapidity coverage from $|y|<0.5$ to $|y|<1$ [\onlinecite{PhysRevD.94.114029}]. The vertical bars represent the statistical uncertainties, while the brackets and transparent boxes represent the systematic uncertainties that are uncorrelated and correlated across the $p_\text{T}$ bins, respectively.
        }
    \end{figure}

\begin{figure}[h!]
        \includegraphics[width = 8.2 cm]{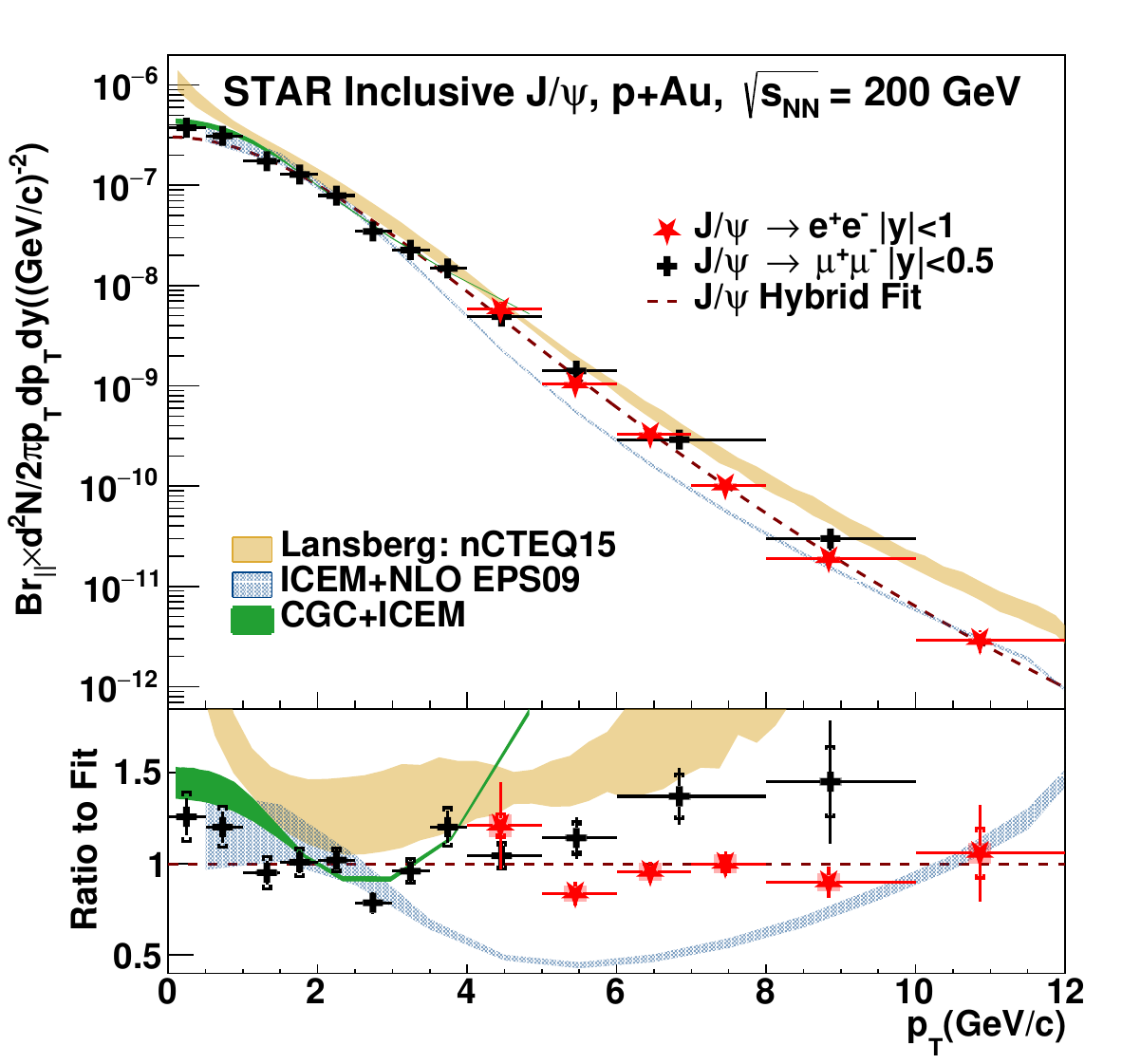}
        \caption{
            \label{fig:fig3_pAu_yield}  Inclusive $J/\psi\rightarrow e^{+}e^{-}$ invariant yield as a function of $p_\text{T}$ in $p+\text{Au}$ collisions at $\sqrt{s_\text{NN}} = 200\text{~GeV}$, compared to the STAR $J/\psi\rightarrow \mu^{+}\mu^{-}$ measurement for $|y| < 0.5$ at the same $\sqrt{s_\text{NN}}$, and to various model calculations. The representation of uncertainties is identical to Fig. \ref{fig:fig2_pp_crosssec}.
        }
    \end{figure}

\begin{figure}[h!]
        \subfloat[\label{fig:fig4.a}]{
            \includegraphics[width = 8.2 cm]{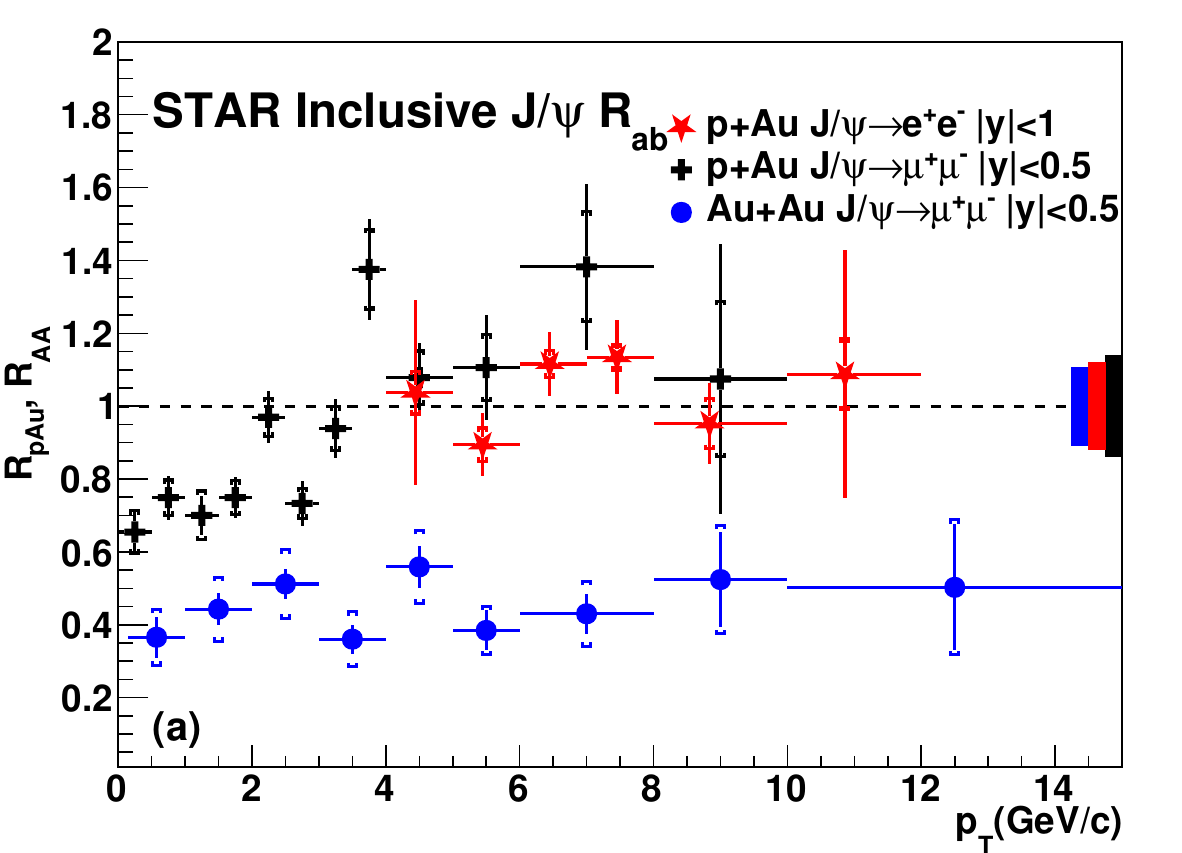}
        }\\
        \subfloat[\label{fig:fig4.b}]{
            \includegraphics[width = 8.2 cm]{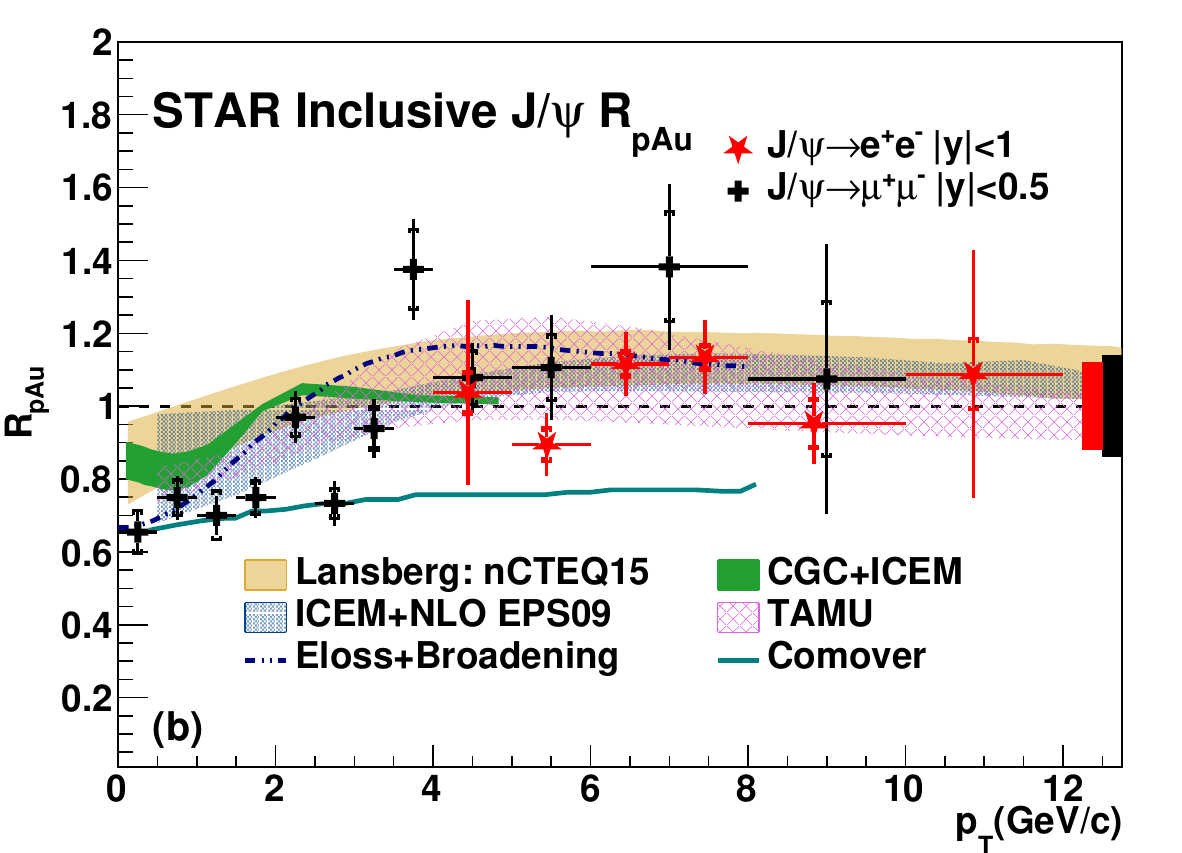}
        }
        \caption{
            \label{fig:fig4_RpAu_data} The $R_{p\text{Au}}$ of inclusive $J/\psi\rightarrow e^{+}e^{-}$  and $J/\psi\rightarrow \mu^{+}\mu^{-}$ at $\sqrt{s_\text{NN}}=200\text{~GeV}$. (a) Compared to the $R_\text{AA}$ in 0--20\% central $\text{Au}+\text{Au}$ collisions at the same $\sqrt{s_\text{NN}}$. (b) Compared to various model calculations. 
            The vertical bars represent the statistical uncertainties, while the brackets and solid boxes of the corresponding color around unity represent the $p_\text{T}$-uncorrelated and $p_\text{T}$-correlated systematic uncertainties, respectively.
        }
    \end{figure}

    \indent Each individual measurement using different years for the $p+p$ yield, as well as the combined one, is converted into a cross section. The expectation of $p_\text{T}$ in each bin is shifted away from the bin center. The acceptance- and efficiency-corrected yield as a function of $p_\text{T}$ is fit to an empirical function using an iterative Bayesian approach, with priors set to the bin centers. In each iteration, the $p_\text{T}$ in each bin is unfolded by searching the $p_\text{T}$ at which the function value is equal to the average in that bin. The empirical function takes the form of:
    \begin{equation}
        f\left(p_\text{T}\right) \propto p_\text{T}\cdot \left(1+\left(\frac{p_\text{T}}{p_\text{T,0}}\right)^{2}\right)^{-n},
        \label{eq:yield_fit}
    \end{equation}
    where $p_\text{T,0}$ and $n$ are free parameters. Figure \ref{fig:fig2_pp_crosssec}\subref{fig:fig2.a} shows the individual measurements of the differential cross section of inclusive $J/\psi$ times the dielectron decay branching ratio in $p+p$ collisions at $\sqrt{s} = 200 \text{~GeV}$  within $|y|<1$, along with the combined result. A function describing the differential cross section of the combined result is derived from its aforementioned fitted function, taking the form of a normalization constant $N$ multiplied by the right-hand side of Equation \ref{eq:yield_fit} with the $p_\text{T}$ term eliminated. It serves as the baseline for the ratios. The parameters are: $N = (4.5\pm0.9) \text{~nb}\cdot\left(\text{GeV/}c\right)^{-2}$, $p_\text{T,0} = (3.4\pm0.4) \text{~GeV/}c$, $n = 4.9\pm0.5$. The bottom panel shows the ratio to the combined fit function. The combined data points are not shown in the bottom panel for visual clarity. The three individual measurements via dielectron decays are consistent within uncertainties, and the combination has improved the precision, as discussed previously.\\
    \indent Figure \ref{fig:fig2_pp_crosssec}\subref{fig:fig2.b} overlays the combined dielectron measurement with the dimuon measurement [\onlinecite{2022136865}] and various model calculations for $|y| < 0.5$, with correction factors applied to convert the measurements of different rapidity coverages into a common range of $|y| < 1$. The correction factors are based on calculations of the Improved Color Evaporation Model (ICEM) [\onlinecite{PhysRevD.94.114029}] as a function of $p_\text{T}$. The results from the dimuon and dielectron decay channels are consistent within uncertainties. They are complementary, providing more precise measurements across different $p_\text{T}$ ranges. The Lansberg model calculation [\onlinecite{Lansberg2016, PhysRevLett.121.052004, SHAO20132562, SHAO2016238}] uses the CT14 proton PDF at Next-to-Leading Order (NLO) [\onlinecite{PhysRevD.93.033006}], with its uncertainties primarily governed by the factorization scale. In the ICEM [\onlinecite{PhysRevD.94.114029}], the transition probability from $c\bar{c}$ pairs to $J/\psi$ is determined by fitting to previously published $J/\psi$ measurements, with the uncertainties originating from different assigned values for the charm quark mass ($m_\text{c}$), factorization and renormalization scales. The ICEM calculation only includes directly produced $J/\psi$ plus those produced in the decays of excited charmonium states. Labeled as ``ICEM+FONLL'', the contribution to $J/\psi$ from $b$-hadrons is taken into account on top of ICEM. Such a contribution ($b$-hadron feeddown) is calculated at the Fixed Order plus Next-to-Leading Logarithms (FONLL) [\onlinecite{Matteo_Cacciari_1998, Matteo_Cacciari_2001}] level and increases from less than 1\% for $p_\text{T}<1$~GeV/$c$ to slightly over 10\% at $p_\text{T}=14$~GeV/$c$. The associated uncertainty is the quadrature sum of the uncertainties of the FONLL and ICEM contributions. Two Next-to-Leading-Order Non-Relativistic QCD (NLO NRQCD) calculations, fit to different experimental data in different $p_\text{T}$ ranges, are included, labeled as ``NRQCD A'' [\onlinecite{PhysRevD.84.114001}] and ``NRQCD B'' [\onlinecite{PhysRevLett.108.172002}]. The calculation labeled as ``CGC+ICEM'' [\onlinecite{PhysRevC.97.014909}] uses the CGC framework to obtain the $c\bar{c}$ production cross section and the ICEM model to describe hadronization. The charm mass is set to be 1.3 GeV/$c^{2}$, and the associated uncertainty is negligible. Both the ``CGC+ICEM'' and Lansberg calculations are tuned to the previous inclusive $J/\psi$ cross-section measurement at 200 GeV [\onlinecite{PhysRevD.85.092004}], allowing a direct comparison with this measurement. The ICEM and ``CGC+ICEM'' results agree with the data within uncertainties up to $p_\text{T}$ of about 3.5 GeV/$c$, after which they deviate from the data. The ICEM tends to return to the data at high $p_\text{T}$. The Lansberg calculation is consistently above the data. The new measurement could improve the parameterization of these non-NRQCD models, especially for $p_\text{T}>4\text{~GeV/}c$. Both NRQCD calculations have large uncertainties, and thus could not be constrained by the new measurement meaningfully.\\
    \indent The yield in $p+\text{Au}$ collisions is converted into the invariant yield using a similar approach as for the $p+p$ cross section. The $J/\psi$ invariant yield in $p+\text{Au}$ collisions measured via dielectron decay channel is shown in Figure \ref{fig:fig3_pAu_yield}, overlaid with the similar measurement via dimuon channel and various model calculations. The parameters of the baseline function used for computing ratios are $N = (3.0\pm0.3)\times 10^{-7}\text{~}\left(\text{GeV/}c\right)^{-2}$, $p_\text{T,0} = (4.4\pm0.5) \text{~GeV/}c$, $n = 6.0\pm0.9$. It is a hybrid fit using data points at $p_\text{T}<4\text{~GeV/}c$ from the dimuon result and at $p_\text{T}>4\text{~GeV/}c$ from the dielectron result. The new dielectron measurement provides significantly better precision at mid-to-high $p_\text{T}$.\\
    \indent The Lansberg calculation [\onlinecite{Lansberg2016, PhysRevLett.121.052004, SHAO20132562, SHAO2016238}] utilizes the nCTEQ15 nPDF at NLO [\onlinecite{PhysRevD.93.085037}], which is constrained by the $J/\psi$ measurements at the LHC. The associated uncertainty accounts for the nPDF uncertainty at the 68\% confidence level and variations in the factorization scale. The EPPS16 nPDF at NLO [\onlinecite{Eskola2017}] yields a very similar result (not shown) within the same framework. The ICEM [\onlinecite{PhysRevD.94.114029}] utilizes the NLO EPS09 nPDF [\onlinecite{K.J._Eskola_2009}], with an associated uncertainty arising from the nPDF uncertainties. The CGC+ICEM [\onlinecite{PhysRevC.97.014909}] directly calculates the $c\bar{c}$ production cross section in $p+\text{Au}$ collisions based on the CGC framework with uncertainties mostly coming from the variation of the average momentum of soft color exchanges and the scale factor between the saturation scales for a proton and a gold nucleus. The contribution of $J/\psi$ from $b$-hadron decays is not included in any of the model calculations for $p+\text{Au}$ collisions, yet the level of agreement between data and model calculations is similar to that seen for $p+p$ collisions.\\
    \indent Figure \ref{fig:fig4_RpAu_data}\subref{fig:fig4.a} illustrates the $R_{p\text{Au}}$ measured via the dielectron decay channel and via the dimuon channel, as well as the $R_\text{AA}$ in $\text{Au}+\text{Au}$ collisions at $\sqrt{s_\text{NN}} = 200\text{~GeV}$ measured by STAR [\onlinecite{2019134917}]. The dielectron measurement significantly improves the precision at mid-to-high $p_\text{T}$ compared to the published dimuon measurement. In the covered $p_\text{T}$ range, the $R_{p\text{Au}}$ is consistent with unity within uncertainties, indicating negligible net modification due to CNM effects. The result supports the conclusion that the suppression of $J/\psi$ in $\text{Au}+\text{Au}$ collision is dominated by the effects of the hot medium.\\
    \indent Figure \ref{fig:fig4_RpAu_data}\subref{fig:fig4.b} compares the $R_{p\text{Au}}$ measured via the dielectron and dimuon decay channels with various model calculations. The Lansberg and ICEM calculations include only nCTEQ15 [\onlinecite{PhysRevD.93.085037}] parameterizations and nPDF effects based on EPS09 [\onlinecite{K.J._Eskola_2009}], respectively. In the model labeled as ``ELoss+Broadening'' [\onlinecite{Arleo2013}], radiative energy loss and $p_\text{T}$ broadening arise from interactions between fast-moving color-octet $c\bar{c}$ pairs in the nucleus rest frame and the cold medium. The $p_\text{T}$ broadening is responsible for the $J/\psi$ enhancement above 2.5 GeV/$c$. ``ICEM+CGC'' has the same source as in the $p+p$ and $p+\text{Au}$ yield results. The TAMU model [\onlinecite{DU2015147, Du2019}] extends the transport model for heavy-ion collisions to $p+\text{Au}$ collisions, using the NLO EPS09 nPDF [\onlinecite{K.J._Eskola_2009}], in which the $J/\psi$ yield in $p+\text{Au}$ collisions is modified by the short-lived hot medium via both dissociation and recombination. Uncertainties arise from the nPDF, the variation in the broadening parameter used to incorporate the Cronin effect, and the formation times for both $J/\psi$ and the QGP. The model labeled as ``comover'' [\onlinecite{FERREIRO201598}], shown as a solid line, introduces the breakup of $J/\psi$ mesons through interactions with final state particles traveling along with them. The nPDF effect based on leading-order EPS09 parameterization is included in the comover model.\\
    \indent With the exception of the comover model, all other calculations are consistent with the dielectron measurement at unity for $p_\text{T}>4\text{~GeV/}c$. These models predict little modification to $J/\psi$ production and are compatible with each other. Within current experimental and theoretical uncertainties, the dielectron measurement cannot discriminate among different CNM effect mechanisms. A similar measurement in other kinematic regions, e.g., forward/backward rapidity, could be more helpful. The comover interaction is formulated $p_\text{T}$ independently, thus dominated by the low $p_\text{T}$ region, leading to an underprediction of $R_{p\text{A}}$ for $p_\text{T}>3.5\text{~GeV/}c$. This provides strong constraints on future studies of $p_\text{T}$ dependence in the comover model.
\section{Summary}
    \indent The differential cross section and invariant yield of inclusive $J/\psi$ are measured via the dielectron decay channel within $|y| < 1$ and $4 < p_\text{T} < 12\text{~GeV}$/$c$ in $p+p$ and $p+\text{Au}$ collisions at $\sqrt{s_\text{NN}} = 200\text{~GeV}$, using data taken by STAR in 2015. The corresponding $R_{p\text{Au}}$ is calculated.\\
    \indent Compared to the published measurements of the $p+p$ cross section and $p+\text{Au}$ invariant yield, this analysis has significantly smaller uncertainty at mid-to-high $p_\text{T}$. A combined inclusive $J/\psi$ cross section in $p+p$ with higher precision, including the results of this analysis and two earlier STAR papers [\onlinecite{201355, 201887}] is reported.\\
    \indent Compared to the STAR inclusive $J/\psi\rightarrow\mu^{+}\mu^{-}$ $R_{p\text{Au}}$ result for $|y|<0.5$ [\onlinecite{2022136865}], the new measurement has noticeably smaller uncertainties at mid-to-high $p_\text{T}$, which affirms the physics conclusion. Measurements utilizing both dielectron and dimuon decay channels in 2015 data have significantly better precision compared to the published measurement of $R_\text{dAu}$ [\onlinecite{2022136865, PhysRevC.93.064904}]. The improvement is partially due to the identity of trigger setup, detector configuration, and analysis procedure between the data taken in $p+p$ and $p+\text{Au}$ collisions in the same year. \\
    \indent The $R_{p\text{Au}}$ result is consistent with calculations that have taken nPDF effects into account, with the exception that the comover model calculation has a discrepancy at mid-to-high $p_\text{T}$. Meanwhile, the individual invariant yields in the different collision systems exhibit less satisfactory agreement with model calculations, since, compared with $R_{p\text{Au}}$, each individual yield is more sensitive to possible systematic biases in the theoretical models. These measurements could provide strong constraints on models, improving understanding of the quarkonia production mechanism and the CNM effects for $J/\psi$ in ultra-relativistic collisions at RHIC energies.
\begin{acknowledgments}
    \indent We thank the RHIC Operations Group and SDCC at BNL, the NERSC Center at LBNL, and the Open Science Grid consortium for providing resources and support. This work was supported in part by the Office of Nuclear Physics within the U.S. DOE Office of Science, the U.S. National Science Foundation, National Natural Science Foundation of China, Chinese Academy of Science, the Ministry of Science and Technology of China and the Chinese Ministry of Education, NSTC Taipei, the National Research Foundation of Korea, Czech Science Foundation and Ministry of Education, Youth and Sports of the Czech Republic, Hungarian National Research, Development and Innovation Office, New National Excellency Programme of the Hungarian Ministry of Human Capacities, Department of Atomic Energy and Department of Science and Technology of the Government of India, the National Science Centre and WUT ID-UB of Poland, the Ministry of Science, Education and Sports of the Republic of Croatia, German Bundesministerium f\"ur Bildung, Wissenschaft, Forschung and Technologie (BMBF), Helmholtz Association, Ministry of Education, Culture, Sports, Science, and Technology (MEXT), and Japan Society for the Promotion of Science (JSPS).
\end{acknowledgments}
\appendix
\section{PYTHIA 8 Settings for STAR HF Tune}
    \label{apdx:apdx1_HF_TUNE}
    \indent The HF Tune is a set of parameters tuned to match the existing measurement of $J/\psi$ and non-photonic electron $p_\text{T}$ distribution. The following are based on the default settings of PYTHIA 8.1.62 [\onlinecite{Sjostrand:2006za, SJOSTRAND2008852}] and LHAPDF 6.1.4 [\onlinecite{Buckley2015}]\\
    \begin{itemize}
        \item SigmaProcess:renormScale2 = 3
        \item SigmaProcess:factorScale2 = 3
        \item SigmaProcess:renormMultFac = 2
        \item SigmaProcess:factorMultFac = 2
        \item PDF:useLHAPDF = on
        \item PDF:LHAPDFset = MRSTMCal.LHgrid
        \item PDF:extrapolateLHAPDF = on
        \item PartonLevel:MI = on
        \item PartonLevel:ISR = on
        \item BeamRemnants:primordialKT = on
        \item PartonLevel:FSR = on
        \item StringFlav:mesonCvector = 1.5
        \item StringFlav:mesonBvector = 3
        \item 4:m0 = 1.43
        \item 5:m0 = 4.30
    \end{itemize}
\section{HIJING Settings}
    \label{apdx:apdx2_HIJING}
    \indent The following settings are based on official default HIJING settings. They are configured for MB events (and for $D^{0}$ events in the parentheses if values differ, respectively).\\
    \begin{itemize}
        \item Impact parameter = 15.0 fm
        \item Jet quenching = off               
        \item B production = on (off)
        \item Charm production = off (on)
        \item Maximum number of hard scatterings = 10
        \item Track all particles = on  
    \end{itemize}

    

\bibliography{PaperDraft}

@PREAMBLE{
 "\providecommand{\noopsort}[1]{}" 
 # "\providecommand{\singleletter}[1]{#1}%" 
}

@article{MATSUI1986416,
    title = {J/ψ suppression by quark-gluon plasma formation},
    journal = {Physics Letters B},
    volume = {178},
    number = {4},
    pages = {416-422},
    year = {1986},
    issn = {0370-2693},
    doi = {https://doi.org/10.1016/0370-2693(86)91404-8},
    url = {https://www.sciencedirect.com/science/article/pii/0370269386914048},
    author = {T. Matsui and H. Satz}
}

@article{PhysRevLett.99.211602,
  title = {Color Screening Melts Quarkonium},
  author = {M\'ocsy, \'Agnes and Petreczky, P\'eter},
  journal = {Phys. Rev. Lett.},
  volume = {99},
  issue = {21},
  pages = {211602},
  numpages = {4},
  year = {2007},
  month = {Nov},
  publisher = {American Physical Society},
  doi = {10.1103/PhysRevLett.99.211602},
  url = {https://link.aps.org/doi/10.1103/PhysRevLett.99.211602}
}

@Article{Andronic2016,
    author={Andronic, A. and et al.},
    title={Heavy-flavour and quarkonium production in the LHC era: from proton--proton to heavy-ion collisions},
    journal={The European Physical Journal C},
    year={2016},
    month={Feb},
    day={29},
    volume={76},
    number={3},
    pages={107},
    issn={1434-6052},
    doi={10.1140/epjc/s10052-015-3819-5},
    url={https://doi.org/10.1140/epjc/s10052-015-3819-5}
}

@article{ABREU200185,
    title = {Transverse momentum distributions of J/ψ, ψ′, Drell–Yan and continuum dimuons produced in Pb–Pb interactions at the SPS},
    journal = {Physics Letters B},
    volume = {499},
    number = {1},
    pages = {85-96},
    year = {2001},
    issn = {0370-2693},
    doi = {https://doi.org/10.1016/S0370-2693(01)00019-3},
    url = {https://www.sciencedirect.com/science/article/pii/S0370269301000193},
    author = {M.C Abreu and et al.}
}

@article{PhysRevLett.98.232301,
  title = {$J/\ensuremath{\psi}$ Production versus Centrality, Transverse Momentum, and Rapidity in $\mathrm{Au}+\mathrm{Au}$ Collisions at $\sqrt{{s}_{NN}}=200\text{ }\text{ }\mathrm{GeV}$},
  author = {Adare, A. and et al.},
  collaboration = {PHENIX Collaboration},
  journal = {Phys. Rev. Lett.},
  volume = {98},
  issue = {23},
  pages = {232301},
  numpages = {6},
  year = {2007},
  month = {Jun},
  publisher = {American Physical Society},
  doi = {10.1103/PhysRevLett.98.232301},
  url = {https://link.aps.org/doi/10.1103/PhysRevLett.98.232301}
}

@article{2011294,
    title = {Measurement of the centrality dependence of J/ψ yields and observation of Z production in lead–lead collisions with the ATLAS detector at the LHC},
    journal = {Physics Letters B},
    volume = {697},
    number = {4},
    pages = {294-312},
    year = {2011},
    issn = {0370-2693},
    doi = {https://doi.org/10.1016/j.physletb.2011.02.006},
    url = {https://www.sciencedirect.com/science/article/pii/S0370269311001389},
    author = {G. Aad and et al.}
}

@Article{Chatrchyan2012,
    author={Chatrchyan, S. and et al.},
    title={Suppression of non-prompt J/$\psi$, prompt J/$\psi$, and {\$} {\backslash}Upsilon {\$}(1S) in PbPb collisions at {\$} {\backslash}sqrt {\{}{\{}{\{}s{\_}{\{}{\backslash}text{\{}NN{\}}{\}}{\}}{\}}{\}} = 2.76 {\$}TeV},
    collaboration = {CMS Collaboration},
    journal={Journal of High Energy Physics},
    year={2012},
    month={May},
    day={14},
    volume={2012},
    number={5},
    pages={63},
    issn={1029-8479},
    doi={10.1007/JHEP05(2012)063},
    url={https://doi.org/10.1007/JHEP05(2012)063}
}

@article{PhysRevLett.109.072301,
  title = {$J/\ensuremath{\psi}$ Suppression at Forward Rapidity in Pb-Pb Collisions at $\sqrt{{s}_{\mathrm{NN}}}=2.76\text{ }\text{ }\mathrm{TeV}$},
  author = {Abelev, B. and et al.},
  collaboration = {ALICE Collaboration},
  journal = {Phys. Rev. Lett.},
  volume = {109},
  issue = {7},
  pages = {072301},
  numpages = {11},
  year = {2012},
  month = {Aug},
  publisher = {American Physical Society},
  doi = {10.1103/PhysRevLett.109.072301},
  url = {https://link.aps.org/doi/10.1103/PhysRevLett.109.072301}
}

@article{PhysRevC.87.034904,
  title = {Transverse-momentum dependence of the $J/\ensuremath{\psi}$ nuclear modification in $d$+Au collisions at $\sqrt{{s}_{NN}}=200$ GeV},
  author = {Adare, A. and et al.},
  collaboration = {PHENIX Collaboration},
  journal = {Phys. Rev. C},
  volume = {87},
  issue = {3},
  pages = {034904},
  numpages = {22},
  year = {2013},
  month = {Mar},
  publisher = {American Physical Society},
  doi = {10.1103/PhysRevC.87.034904},
  url = {https://link.aps.org/doi/10.1103/PhysRevC.87.034904}
}

@article{PhysRevC.102.014902,
  title = {Measurement of $J/\ensuremath{\psi}$ at forward and backward rapidity in $p+p$, $p+\mathrm{Al}$, $p+\mathrm{Au}$, and $^{3}\mathrm{He}+\mathrm{Au}$ collisions at $\sqrt{{s}_{NN}}=200$ GeV},
  author = {Acharya, U. and et al.},
  collaboration = {PHENIX Collaboration},
  journal = {Phys. Rev. C},
  volume = {102},
  issue = {1},
  pages = {014902},
  numpages = {23},
  year = {2020},
  month = {Jul},
  publisher = {American Physical Society},
  doi = {10.1103/PhysRevC.102.014902},
  url = {https://link.aps.org/doi/10.1103/PhysRevC.102.014902}
}

@article{2022136865,
    title = {Measurement of cold nuclear matter effects for inclusive J/ψ in p+Au collisions at sNN=200 GeV},
    collaboration = {STAR Collaboration},
    journal = {Physics Letters B},
    volume = {825},
    pages = {136865},
    year = {2022},
    issn = {0370-2693},
    doi = {https://doi.org/10.1016/j.physletb.2021.136865},
    url = {https://www.sciencedirect.com/science/article/pii/S0370269321008054},
    author = {M.S. Abdallah and et al.}
}

@Article{Acharya2018,
    author={Acharya, S. and et al.},
    collaboration = {ALICE Collaboration},
    title={Inclusive J/$\psi$ production at forward and backward rapidity in p-Pb collisions at {\$}{\$} {\backslash}sqrt{\{}s{\_}{\{}{\backslash}mathrm{\{}NN{\}}{\}}{\}}=8.16 {\$}{\$}TeV},
    journal={Journal of High Energy Physics},
    year={2018},
    month={Jul},
    day={25},
    volume={2018},
    number={7},
    pages={160},
    issn={1029-8479},
    doi={10.1007/JHEP07(2018)160},
    url={https://doi.org/10.1007/JHEP07(2018)160}
}

@article{PhysRevD.93.085037,
  title = {nCTEQ15: Global analysis of nuclear parton distributions with uncertainties in the CTEQ framework},
  author = {Kova\ifmmode \check{r}\else \v{r}\fi{}\'{\i}k, K. and Kusina, A. and Je\ifmmode \check{z}\else \v{z}\fi{}o, T. and Clark, D. B. and Keppel, C. and Lyonnet, F. and Morf\'{\i}n, J. G. and Olness, F. I. and Owens, J. F. and Schienbein, I. and Yu, J. Y.},
  journal = {Phys. Rev. D},
  volume = {93},
  issue = {8},
  pages = {085037},
  numpages = {34},
  year = {2016},
  month = {Apr},
  publisher = {American Physical Society},
  doi = {10.1103/PhysRevD.93.085037},
  url = {https://link.aps.org/doi/10.1103/PhysRevD.93.085037}
}

@Article{Eskola2017,
    author={Eskola, Kari J.
    and Paakkinen, Petja
    and Paukkunen, Hannu
    and Salgado, Carlos A.},
    title={EPPS16: nuclear parton distributions with LHC data},
    journal={The European Physical Journal C},
    year={2017},
    month={Mar},
    day={16},
    volume={77},
    number={3},
    pages={163},
    issn={1434-6052},
    doi={10.1140/epjc/s10052-017-4725-9},
    url={https://doi.org/10.1140/epjc/s10052-017-4725-9}
}

@article{CGC2010,
    author={
        Gelis, Francois and
        Iancu, Edmond and 
        Jalilian-Marian, Jamal and
        Venugopalan, Raju
    },
    title={The Color Glass Condensate},
    journal={Annual Review of Nuclear and Particle Science},
    year={2010},
    month={Nov},
    day={23},
    volume={60},
    pages={463-489},
    url={https://doi.org/10.1146/annurev.nucl.010909.083629}
}

@Article{Arleo2013,
    author={Arleo, Fran{\c{c}}ois
    and Kolevatov, Rodion
    and Peign{\'e}, St{\'e}phane
    and Rustamova, Maryam},
    title={Centrality and p⊥ dependence of J/$\psi$ suppression in proton-nucleus collisions from parton energy loss},
    journal={Journal of High Energy Physics},
    year={2013},
    month={May},
    day={29},
    volume={2013},
    number={5},
    pages={155},
    issn={1029-8479},
    doi={10.1007/JHEP05(2013)155},
    url={https://doi.org/10.1007/JHEP05(2013)155}
}

@article{VOGT1999197,
    title = {J/ψ production and suppression11This work was supported in part by the Director, Office of Energy Research, Division of Nuclear Physics of the Office of High Energy and Nuclear Physics of the U.S. Department of Energy under Contract Number DE-AC03-76SF00098.},
    journal = {Physics Reports},
    volume = {310},
    number = {4},
    pages = {197-260},
    year = {1999},
    issn = {0370-1573},
    doi = {https://doi.org/10.1016/S0370-1573(98)00074-X},
    url = {https://www.sciencedirect.com/science/article/pii/S037015739800074X},
    author = {R. Vogt}
}

@article{FERREIRO201598,
    title = {Excited charmonium suppression in proton–nucleus collisions as a consequence of comovers},
    journal = {Physics Letters B},
    volume = {749},
    pages = {98-103},
    year = {2015},
    issn = {0370-2693},
    doi = {https://doi.org/10.1016/j.physletb.2015.07.066},
    url = {https://www.sciencedirect.com/science/article/pii/S0370269315005766},
    author = {E.G. Ferreiro}
}

@article{201355,
    title = {J/ψ production at high transverse momenta in p+p and Au + Au collisions at sNN=200 GeV},
    journal = {Physics Letters B},
    volume = {722},
    number = {1},
    pages = {55-62},
    year = {2013},
    issn = {0370-2693},
    doi = {https://doi.org/10.1016/j.physletb.2013.04.010},
    url = {https://www.sciencedirect.com/science/article/pii/S037026931300289X},
    author = {L. Adamczyk and et al.},
    collaboration={STAR Collaboration}
}

@article{201887,
    title = {J/ψ production cross section and its dependence on charged-particle multiplicity in p + p collisions at s=200 GeV},
    journal = {Physics Letters B},
    volume = {786},
    pages = {87-93},
    year = {2018},
    issn = {0370-2693},
    doi = {https://doi.org/10.1016/j.physletb.2018.09.029},
    url = {https://www.sciencedirect.com/science/article/pii/S0370269318307305},
    author = {J. Adam and et al.},
    collaboration={STAR Collaboration}
}

@article{DU2015147,
    title = {Sequential regeneration of charmonia in heavy-ion collisions},
    journal = {Nuclear Physics A},
    volume = {943},
    pages = {147-158},
    year = {2015},
    issn = {0375-9474},
    doi = {https://doi.org/10.1016/j.nuclphysa.2015.09.006},
    url = {https://www.sciencedirect.com/science/article/pii/S0375947415002055},
    author = {Xiaojian Du and Ralf Rapp}
}

@Article{Du2019,
    author={Du, Xiaojian
    and Rapp, Ralf},
    title={In-medium charmonium production in proton-nucleus collisions},
    journal={Journal of High Energy Physics},
    year={2019},
    month={Mar},
    day={05},
    volume={2019},
    number={3},
    pages={15},
    issn={1029-8479},
    doi={10.1007/JHEP03(2019)015},
    url={https://doi.org/10.1007/JHEP03(2019)015}
}

@article{PhysRevD.74.074007,
  title = {Nonrelativistic QCD factorization and the velocity dependence of NNLO poles in heavy quarkonium production},
  author = {Nayak, Gouranga C. and Qiu, Jian-Wei and Sterman, George},
  journal = {Phys. Rev. D},
  volume = {74},
  issue = {7},
  pages = {074007},
  numpages = {13},
  year = {2006},
  month = {Oct},
  publisher = {American Physical Society},
  doi = {10.1103/PhysRevD.74.074007},
  url = {https://link.aps.org/doi/10.1103/PhysRevD.74.074007}
}

@article{PhysRevLett.91.172302,
  title = {Transverse-Momentum and Collision-Energy Dependence of High-${p}_{T}$ Hadron Suppression in $\mathrm{A}\mathrm{u}+\mathrm{A}\mathrm{u}$ Collisions at Ultrarelativistic Energies},
  author = {Adams, J. and et al.},
  collaboration = {STAR Collaboration},
  journal = {Phys. Rev. Lett.},
  volume = {91},
  issue = {17},
  pages = {172302},
  numpages = {6},
  year = {2003},
  month = {Oct},
  publisher = {American Physical Society},
  doi = {10.1103/PhysRevLett.91.172302},
  url = {https://link.aps.org/doi/10.1103/PhysRevLett.91.172302}
}

@Article{Buckley2015,
    author={Buckley, Andy
    and Ferrando, James
    and Lloyd, Stephen
    and Nordstr{\"o}m, Karl
    and Page, Ben
    and R{\"u}fenacht, Martin
    and Sch{\"o}nherr, Marek
    and Watt, Graeme},
    title={LHAPDF6: parton density access in the LHC precision era},
    journal={The European Physical Journal C},
    year={2015},
    month={Mar},
    day={20},
    volume={75},
    number={3},
    pages={132},
    issn={1434-6052},
    doi={10.1140/epjc/s10052-015-3318-8},
    url={https://doi.org/10.1140/epjc/s10052-015-3318-8}
}

@Article{Corke2011,
    author={Corke, Richard
    and Sj{\"o}strand, Torbj{\"o}rn},
    title={Multiparton interactions with an x-dependent proton size},
    journal={Journal of High Energy Physics},
    year={2011},
    month={May},
    day={02},
    volume={2011},
    number={5},
    pages={9},
    issn={1029-8479},
    doi={10.1007/JHEP05(2011)009},
    url={https://doi.org/10.1007/JHEP05(2011)009}
}

@article{PhysRevD.93.033006,
  title = {New parton distribution functions from a global analysis of quantum chromodynamics},
  author = {Dulat, Sayipjamal and Hou, Tie-Jiun and Gao, Jun and Guzzi, Marco and Huston, Joey and Nadolsky, Pavel and Pumplin, Jon and Schmidt, Carl and Stump, Daniel and Yuan, C.-P.},
  journal = {Phys. Rev. D},
  volume = {93},
  issue = {3},
  pages = {033006},
  numpages = {39},
  year = {2016},
  month = {Feb},
  publisher = {American Physical Society},
  doi = {10.1103/PhysRevD.93.033006},
  url = {https://link.aps.org/doi/10.1103/PhysRevD.93.033006}
}

@article{PhysRevD.94.114029,
  title = {Quarkonium production in an improved color evaporation model},
  author = {Ma, Yan-Qing and Vogt, Ramona},
  journal = {Phys. Rev. D},
  volume = {94},
  issue = {11},
  pages = {114029},
  numpages = {6},
  year = {2016},
  month = {Dec},
  publisher = {American Physical Society},
  doi = {10.1103/PhysRevD.94.114029},
  url = {https://link.aps.org/doi/10.1103/PhysRevD.94.114029}
}

@article{Matteo_Cacciari_1998,
    doi = {10.1088/1126-6708/1998/05/007},
    url = {https://dx.doi.org/10.1088/1126-6708/1998/05/007},
    year = {1998},
    month = {jun},
    publisher = {},
    volume = {1998},
    number = {05},
    pages = {007},
    author = {Matteo Cacciari and  Mario Greco and  Paolo Nason},
    title = {The  pT spectrum in heavy-flavour hadroproduction},
    journal = {Journal of High Energy Physics}
}

@article{Matteo_Cacciari_2001,
    doi = {10.1088/1126-6708/2001/03/006},
    url = {https://dx.doi.org/10.1088/1126-6708/2001/03/006},
    year = {2001},
    month = {mar},
    publisher = {},
    volume = {2001},
    number = {03},
    pages = {006},
    author = {Matteo Cacciari and  Stefano Frixione and  Paolo Nason},
    title = {The  pT spectrum in heavy-flavour photoproduction},
    journal = {Journal of High Energy Physics}
}

@article{PhysRevD.85.092004,
  title = {Ground and excited state charmonium production in $p+p$ collisions at $\sqrt{s}=200\text{ }\text{ }\mathrm{GeV}$},
  author = {Adare, A. and et al.},
  collaboration = {PHENIX Collaboration},
  journal = {Phys. Rev. D},
  volume = {85},
  issue = {9},
  pages = {092004},
  numpages = {27},
  year = {2012},
  month = {May},
  publisher = {American Physical Society},
  doi = {10.1103/PhysRevD.85.092004},
  url = {https://link.aps.org/doi/10.1103/PhysRevD.85.092004}
}

@article{K.J._Eskola_2009,
    doi = {10.1088/1126-6708/2009/04/065},
    url = {https://dx.doi.org/10.1088/1126-6708/2009/04/065},
    year = {2009},
    month = {apr},
    publisher = {},
    volume = {2009},
    number = {04},
    pages = {065},
    author = {K.J. Eskola and  H. Paukkunen and  C.A. Salgado},
    title = {EPS09 — A new generation of NLO and LO nuclear parton distribution functions},
    journal = {Journal of High Energy Physics}
}

@Article{Lansberg2016,
    author={Lansberg, Jean-Philippe and Shao, Hua-Sheng},
    title={Towards an automated tool to evaluate the impact of the nuclear modification of the gluon density on quarkonium, D and B meson production in proton--nucleus collisions},
    journal={The European Physical Journal C},
    year={2016},
    month={Dec},
    day={27},
    volume={77},
    number={1},
    pages={1},
    issn={1434-6052},
    doi={10.1140/epjc/s10052-016-4575-x},
    url={https://doi.org/10.1140/epjc/s10052-016-4575-x}
}

@article{PhysRevLett.121.052004,
  title = {Gluon Shadowing in Heavy-Flavor Production at the LHC},
  author = {Kusina, Aleksander and Lansberg, Jean-Philippe and Schienbein, Ingo and Shao, Hua-Sheng},
  journal = {Phys. Rev. Lett.},
  volume = {121},
  issue = {5},
  pages = {052004},
  numpages = {8},
  year = {2018},
  month = {Aug},
  publisher = {American Physical Society},
  doi = {10.1103/PhysRevLett.121.052004},
  url = {https://link.aps.org/doi/10.1103/PhysRevLett.121.052004}
}

@article{SHAO20132562,
    title = {HELAC-Onia: An automatic matrix element generator for heavy quarkonium physics},
    journal = {Computer Physics Communications},
    volume = {184},
    number = {11},
    pages = {2562-2570},
    year = {2013},
    issn = {0010-4655},
    doi = {https://doi.org/10.1016/j.cpc.2013.05.023},
    url = {https://www.sciencedirect.com/science/article/pii/S0010465513001938},
    author = {Hua-Sheng Shao},
}

@article{SHAO2016238,
    title = {HELAC-Onia 2.0: An upgraded matrix-element and event generator for heavy quarkonium physics},
    journal = {Computer Physics Communications},
    volume = {198},
    pages = {238-259},
    year = {2016},
    issn = {0010-4655},
    doi = {https://doi.org/10.1016/j.cpc.2015.09.011},
    url = {https://www.sciencedirect.com/science/article/pii/S0010465515003495},
    author = {Hua-Sheng Shao},
}

@article{PhysRevC.93.064904,
    title = {$J/\ensuremath{\psi}$ production at low transverse momentum in $p+p$ and $d$ + Au collisions at $\sqrt{{s}_{NN}}=200$ GeV},
    author = {Adamczyk, L. and et al.},
    collaboration = {STAR Collaboration},
    journal = {Phys. Rev. C},
    volume = {93},
    issue = {6},
    pages = {064904},
    numpages = {11},
    year = {2016},
    month = {Jun},
    publisher = {American Physical Society},
    doi = {10.1103/PhysRevC.93.064904},
    url = {https://link.aps.org/doi/10.1103/PhysRevC.93.064904}
}

@article{PhysRevD.100.014008,
    title = {Quarkonium inside the quark-gluon plasma: Diffusion, dissociation, recombination, and energy loss},
    author = {Yao, Xiaojun and M\"uller, Berndt},
    journal = {Phys. Rev. D},
    volume = {100},
    issue = {1},
    pages = {014008},
    numpages = {16},
    year = {2019},
    month = {Jul},
    publisher = {American Physical Society},
    doi = {10.1103/PhysRevD.100.014008},
    url = {https://link.aps.org/doi/10.1103/PhysRevD.100.014008}
}

@article{PhysRevC.53.3051,
    title = {J/\ensuremath{\psi} suppression in an equilibrating parton plasma},
    author = {Xu, Xiao-Ming and Kharzeev, D. and Satz, H. and Wang, Xin-Nian},
    journal = {Phys. Rev. C},
    volume = {53},
    issue = {6},
    pages = {3051--3056},
    numpages = {0},
    year = {1996},
    month = {Jun},
    publisher = {American Physical Society},
    doi = {10.1103/PhysRevC.53.3051},
    url = {https://link.aps.org/doi/10.1103/PhysRevC.53.3051}
}

@article{PhysRevC.87.044905,
    title = {High transverse momentum quarkonium production and dissociation in heavy ion collisions},
    author = {Sharma, Rishi and Vitev, Ivan},
    journal = {Phys. Rev. C},
    volume = {87},
    issue = {4},
    pages = {044905},
    numpages = {23},
    year = {2013},
    month = {Apr},
    publisher = {American Physical Society},
    doi = {10.1103/PhysRevC.87.044905},
    url = {https://link.aps.org/doi/10.1103/PhysRevC.87.044905}
}

@Article{Nisius2014,
    author={Nisius, Richard},
    title={On the combination of correlated estimates of a physics observable},
    journal={The European Physical Journal C},
    year={2014},
    month={Aug},
    day={09},
    volume={74},
    number={8},
    pages={3004},
    issn={1434-6052},
    doi={10.1140/epjc/s10052-014-3004-2},
    url={https://doi.org/10.1140/epjc/s10052-014-3004-2}
}

@article{PhysRevC.97.014909,
    title = {$\ensuremath{\psi}(2S)$ versus $J/\ensuremath{\psi}$ suppression in proton-nucleus collisions from factorization violating soft color exchanges},
    author = {Ma, Yan-Qing and Venugopalan, Raju and Watanabe, Kazuhiro and Zhang, Hong-Fei},
    journal = {Phys. Rev. C},
    volume = {97},
    issue = {1},
    pages = {014909},
    numpages = {13},
    year = {2018},
    month = {Jan},
    publisher = {American Physical Society},
    doi = {10.1103/PhysRevC.97.014909},
    url = {https://link.aps.org/doi/10.1103/PhysRevC.97.014909}
}

@article{PhysRevLett.108.172002,
    title = {$J/\ensuremath{\psi}$ Polarization at the Tevatron and the LHC: Nonrelativistic-QCD Factorization at the Crossroads},
    author = {Butenschoen, Mathias and Kniehl, Bernd A.},
    journal = {Phys. Rev. Lett.},
    volume = {108},
    issue = {17},
    pages = {172002},
    numpages = {5},
    year = {2012},
    month = {Apr},
    publisher = {American Physical Society},
    doi = {10.1103/PhysRevLett.108.172002},
    url = {https://link.aps.org/doi/10.1103/PhysRevLett.108.172002}
}

@article{PhysRevD.84.114001,
    title = {Complete next-to-leading order calculation of the $J/\ensuremath{\psi}$ and ${\ensuremath{\psi}}^{\ensuremath{'}}$ production at hadron colliders},
    author = {Ma, Yan-Qing and Wang, Kai and Chao, Kuang-Ta},
    journal = {Phys. Rev. D},
    volume = {84},
    issue = {11},
    pages = {114001},
    numpages = {10},
    year = {2011},
    month = {Dec},
    publisher = {American Physical Society},
    doi = {10.1103/PhysRevD.84.114001},
    url = {https://link.aps.org/doi/10.1103/PhysRevD.84.114001}
}

@article{ANDERSON2003659,
    title = {The STAR time projection chamber: a unique tool for studying high multiplicity events at RHIC},
    journal = {Nuclear Instruments and Methods in Physics Research Section A: Accelerators, Spectrometers, Detectors and Associated Equipment},
    volume = {499},
    number = {2},
    pages = {659-678},
    year = {2003},
    note = {The Relativistic Heavy Ion Collider Project: RHIC and its Detectors},
    issn = {0168-9002},
    doi = {https://doi.org/10.1016/S0168-9002(02)01964-2},
    url = {https://www.sciencedirect.com/science/article/pii/S0168900202019642},
    author = {M. Anderson and et al.}
}

@article{BEDDO2003725,
    title = {The STAR Barrel Electromagnetic Calorimeter},
    journal = {Nuclear Instruments and Methods in Physics Research Section A: Accelerators, Spectrometers, Detectors and Associated Equipment},
    volume = {499},
    number = {2},
    pages = {725-739},
    year = {2003},
    note = {The Relativistic Heavy Ion Collider Project: RHIC and its Detectors},
    issn = {0168-9002},
    doi = {https://doi.org/10.1016/S0168-9002(02)01970-8},
    url = {https://www.sciencedirect.com/science/article/pii/S0168900202019708},
    author = {M. Beddo and et al.}
}

@article{10.1063/1.2888113,
    author = {Whitten, C. A.},
    collaboration = {STAR Collaboration},
    title = {The Beam‐Beam Counter: A Local Polarimeter at STAR},
    journal = {AIP Conference Proceedings},
    volume = {980},
    number = {1},
    pages = {390-396},
    year = {2008},
    month = {02},
    issn = {0094-243X},
    doi = {10.1063/1.2888113},
    url = {https://doi.org/10.1063/1.2888113},
    eprint = {https://pubs.aip.org/aip/acp/article-pdf/980/1/390/11985098/390\_1\_online.pdf},
}

@article{Smirnov_2017,
    doi = {10.1088/1742-6596/898/4/042058},
    url = {https://dx.doi.org/10.1088/1742-6596/898/4/042058},
    year = {2017},
    month = {oct},
    publisher = {IOP Publishing},
    volume = {898},
    number = {4},
    pages = {042058},
    author = {Smirnov, D. and Lauret, J. and Perevoztchikov, V. and Van Buren, G. and Webb, J.},
    title = {Vertex Reconstruction at STAR: Overview and Performance Evaluation},
    journal = {Journal of Physics: Conference Series}
}

@article{Sjostrand:2006za,
    author = "Sjöstrand, Torbjörn and Mrenna, Stephen and Skands, Peter Z.",
    title = "{PYTHIA 6.4 Physics and Manual}",
    eprint = "hep-ph/0603175",
    archivePrefix = "arXiv",
    reportNumber = "FERMILAB-PUB-06-052-CD-T, LU-TP-06-13",
    doi = "10.1088/1126-6708/2006/05/026",
    journal = "JHEP",
    volume = "05",
    pages = "026",
    year = "2006"
}

@article{SJOSTRAND2008852,
    title = {A brief introduction to PYTHIA 8.1},
    journal = {Computer Physics Communications},
    volume = {178},
    number = {11},
    pages = {852-867},
    year = {2008},
    issn = {0010-4655},
    doi = {https://doi.org/10.1016/j.cpc.2008.01.036},
    url = {https://www.sciencedirect.com/science/article/pii/S0010465508000441},
    author = {Torbjörn Sjöstrand and Stephen Mrenna and Peter Skands}
}

@article{CMS:2017exb,
    author = "Sirunyan, Albert M and et al.",
    collaboration = "CMS",
    title = "{Measurement of prompt and nonprompt $\mathrm{J}/{\psi }$ production in $\mathrm {p}\mathrm {p}$ and $\mathrm {p}\mathrm {Pb}$ collisions at $\sqrt{s_{\mathrm {NN}}} =5.02\,\text {TeV} $}",
    eprint = "1702.01462",
    archivePrefix = "arXiv",
    primaryClass = "nucl-ex",
    reportNumber = "CMS-HIN-14-009, CERN-EP-2017-009",
    doi = "10.1140/epjc/s10052-017-4828-3",
    journal = "Eur. Phys. J. C",
    volume = "77",
    number = "4",
    pages = "269",
    year = "2017"
}

@article{c9wp-5tq3,
  title = {Observation of the Charged-Particle Multiplicity Dependence of ${\ensuremath{\sigma}}_{\ensuremath{\psi}(2S)}/{\ensuremath{\sigma}}_{J/\ensuremath{\psi}}$ in $p$-Pb Collisions at 8.16 TeV},
  author = {Chekhovsky, V. and et al.},
  collaboration = {CMS Collaboration},
  journal = {Phys. Rev. Lett.},
  volume = {135},
  issue = {9},
  pages = {092301},
  numpages = {22},
  year = {2025},
  month = {Aug},
  publisher = {American Physical Society},
  doi = {10.1103/c9wp-5tq3},
  url = {https://link.aps.org/doi/10.1103/c9wp-5tq3}
}

@article{2019134917,
    title = {Measurement of inclusive J/ψ suppression in Au+Au collisions at sNN=200 GeV through the dimuon channel at STAR},
    journal = {Physics Letters B},
    volume = {797},
    pages = {134917},
    year = {2019},
    issn = {0370-2693},
    doi = {https://doi.org/10.1016/j.physletb.2019.134917},
    url = {https://www.sciencedirect.com/science/article/pii/S0370269319306392},
    author = {J. Adam and et al.}
}

@article{PhysRevD.86.072013,
    title = {Measurements of ${D}^{0}$ and ${D}^{\mathbf{*}}$ production in $p\mathbf{+}p$ collisions at $\sqrt{s}\mathbf{=}200\text{ }\text{ }\mathrm{GeV}$},
    author = {Adamczyk, L. and et al.},
    collaboration = {STAR Collaboration},
    journal = {Phys. Rev. D},
    volume = {86},
    issue = {7},
    pages = {072013},
    numpages = {14},
    year = {2012},
    month = {Oct},
    publisher = {American Physical Society},
    doi = {10.1103/PhysRevD.86.072013},
    url = {https://link.aps.org/doi/10.1103/PhysRevD.86.072013}
}

@article{BICHSEL2006154,
    title = {A method to improve tracking and particle identification in TPCs and silicon detectors},
    journal = {Nuclear Instruments and Methods in Physics Research Section A: Accelerators, Spectrometers, Detectors and Associated Equipment},
    volume = {562},
    number = {1},
    pages = {154-197},
    year = {2006},
    issn = {0168-9002},
    doi = {https://doi.org/10.1016/j.nima.2006.03.009},
    url = {https://www.sciencedirect.com/science/article/pii/S0168900206005353},
    author = {Hans Bichsel}
}

@article{LANGE2001152,
    title = {The EvtGen particle decay simulation package},
    journal = {Nuclear Instruments and Methods in Physics Research Section A: Accelerators, Spectrometers, Detectors and Associated Equipment},
    volume = {462},
    number = {1},
    pages = {152-155},
    year = {2001},
    note = {BEAUTY2000, Proceedings of the 7th Int. Conf. on B-Physics at Hadron Machines},
    issn = {0168-9002},
    doi = {https://doi.org/10.1016/S0168-9002(01)00089-4},
    url = {https://www.sciencedirect.com/science/article/pii/S0168900201000894},
    author = {David J. Lange}
}

\end{document}